\documentclass[11pt]{article}
\pdfoutput=1
\usepackage{amsmath,amsfonts,amssymb,color}
\usepackage{cite,url}
\usepackage{graphicx}
\usepackage[utf8]{inputenc} 
\usepackage[bottom]{footmisc}
\usepackage{hyperref}

\allowdisplaybreaks

\def\msk2lam{$m_s\approx 5.5$~TeV, $m_a\approx 4.2$~TeV}

\def\smallNLO{{\rm{\scriptscriptstyle NLO}}}
\def\smallSM{{\rm{\scriptscriptstyle SM}}}
\def\smallBSM{{\rm{\scriptscriptstyle BSM}}}

\def\smallF{{\scriptscriptstyle F}}

\def\smallR{{\scriptscriptstyle R}}

\def\beq{\begin{equation}}
\def\eeq{\end{equation}}
\def\bea{\begin{eqnarray}}
\def\eea{\end{eqnarray}}
\def\nn{\nonumber}
\def\wt{\widetilde}

\def\MM{M^2}

\def\mHq{M^2_H}
\def\mAq{M^2_A}
\def\mHpq{M^2_{H^\pm}}

\def\sq2{\sqrt{2}}

\def\msbar{\overline{\rm MS}}

\def\smallIPI{{\scriptscriptstyle {\rm 1PI}}}

\long\def\symbolfootnote[#1]#2{\begingroup%
\def\thefootnote{\fnsymbol{footnote}}\footnote[#1]{#2}\endgroup}

\makeatletter
\newcommand{\vast}{\bBigg@{3}}
\makeatother

\begin{document}

\begin{titlepage}
\mbox{}
  \vspace*{-4cm}
  \flushright{{\large{\tt COMETA-2025-32}}}
  
\begin{center}

\vspace{2cm}

{\LARGE \bf Two-loop BSM contributions to }\\[3mm]

{\LARGE \bf Higgs pair production in the aligned THDM} 

\vspace{1cm}

{\Large Giuseppe Degrassi$^{\,a,b}$, Ramona Gr\"ober$^{\,c,d}$ and Pietro~Slavich$^{\,e}$}

\vspace*{1cm}

{\sl ${}^a$ 
Dipartimento di Matematica e Fisica, Università di Roma Tre, 

Via della Vasca Navale 84, I-00146 Rome, Italy.}
\vspace*{2mm}\\{\sl ${}^b$ 
  Istituto Nazionale di Fisica Nucleare, Sezione di Roma Tre,

Via della Vasca Navale 84, I-00146 Rome, Italy.}
\vspace*{2mm}\\{\sl ${}^c$
  Dipartimento di Fisica e Astronomia “G. Galilei”, Università di Padova,

  Via F.~Marzolo 8, I-35131 Padua, Italy.}
\vspace*{2mm}\\{\sl ${}^d$
  Istituto Nazionale di Fisica Nucleare, Sezione di Padova,

  Via F.~Marzolo 8, I-35131 Padua, Italy.}
\vspace*{2mm}\\{\sl ${}^e$
   Sorbonne Université, CNRS,
  Laboratoire de Physique Th\'eorique et Hautes Energies, 
 
  4 Place Jussieu, F-75005, Paris, France.}
\end{center}
\symbolfootnote[0]{{\tt e-mail:}}
\symbolfootnote[0]{{\tt giuseppe.degrassi@uniroma3.it}}
\symbolfootnote[0]{{\tt ramona.groeber@pd.infn.it}}
\symbolfootnote[0]{{\tt slavich@lpthe.jussieu.fr}}

\vspace{0.7cm}

\abstract{We study the impact of the two-loop corrections controlled
  by the BSM Higgs couplings on the cross section for the production
  of a pair of SM-like Higgs bosons via gluon fusion in the aligned
  THDM. To this aim, we reassess the two-loop calculation of
  $\lambda_{hhh}$, we compute for the first time the two-loop
  corrections to $\lambda_{hhH}$, and we include the relevant
  corrections to the Higgs-gluon couplings and to the $s$-channel
  propagators entering the $gg\rightarrow hh$ amplitude. We discuss
  the numerical impact of the two-loop BSM contributions, first on the
  individual couplings and then on the prediction for the
  pair-production cross section, in two benchmark scenarios for the
  aligned THDM.}

\vfill

\end{titlepage}


\setcounter{footnote}{0}

\section{Introduction}
\label{sec:intro}

The discovery of a Higgs boson with mass around
$125$~GeV~\cite{CMS:2012qbp, ATLAS:2012yve} and properties compatible
with the predictions of the Standard Model
(SM)~\cite{ParticleDataGroup:2022pth} goes a long way towards
elucidating the mechanism of electroweak (EW) symmetry breaking, but
does not by itself preclude the existence of additional, beyond-the-SM
(BSM) Higgs bosons with masses around or even below the TeV scale,
which could still be discovered in the current or future runs of the
Large Hadron Collider (LHC).

The Two-Higgs-Doublet Model (THDM) is one of the simplest and
best-studied extensions of the SM (for reviews see, e.g.,
refs.~\cite{Gunion:1989we,Aoki:2009ha,Branco:2011iw}). In the
CP-conserving versions of the model, the Higgs sector includes five
physical states: two CP-even scalars, $h$ and $H$, with $M_h<M_H$;
one CP-odd scalar, $A$; and two charged scalars, $H^\pm$. As
discussed, e.g., in ref.~\cite{Gunion:2002zf}, the so-called
``alignment'' condition -- in which one of the CP-even scalars has
SM-like couplings to fermions and gauge bosons -- can be realized
through decoupling, when all of the other Higgs bosons are much
heavier, or without decoupling, when a specific configuration of
parameters in the Lagrangian suppresses the mixing between the SM-like
scalar (typically $h$) and the other CP-even scalar. While the latter
scenario can be viewed as involving a certain amount of fine tuning,
it is often enforced ``from the bottom up'', based on the empirical
observation that the couplings of the $125$-GeV Higgs boson appear to
be essentially SM-like.

Beyond the requirement that they allow for a scalar with mass around
$125$~GeV and SM-like couplings to fermions and gauge bosons, the
parameters of the THDM are subject to a number of experimental
constraints from direct searches for BSM Higgs bosons, EW precision
observables, and flavor physics, as well as theory-driven constraints
from perturbative unitarity, perturbativity, and the stability of the
scalar potential. Nevertheless, quartic Higgs couplings of ${\cal
  O}(1$--$10)$ are still allowed by all constraints (see, e.g.,
refs.~\cite{Arco:2020ucn, Arco:2022xum}).
%
%
Couplings in this range may become the dominant source of radiative
corrections to the THDM predictions for physical observables, up to
the point where one might wonder whether, in any given calculation,
the uncomputed higher-order effects spoil the accuracy of the
prediction.  This has motivated a number of recent studies in which
radiative corrections involving the quartic Higgs couplings of the
THDM were computed at the two-loop level.  In particular, the two-loop
corrections to the $\rho$ parameter were computed in
refs.~\cite{Hessenberger:2016atw, Hessenberger:2022tcx}, various
effects of the two-loop corrections to the scalar mass matrices were
examined in ref.~\cite{Braathen:2017izn}, the two-loop corrections to
the trilinear self-coupling of the SM-like Higgs boson,
$\lambda_{hhh}$, were computed in refs.~\cite{Braathen:2019pxr,
  Braathen:2019zoh}, and the two-loop corrections to the decay width
for the process $h\rightarrow \gamma \gamma$ were computed in
refs.~\cite{Degrassi:2023eii, Aiko:2023nqj}.\footnote{Generic results
for the two-loop corrections to the trilinear and quartic Higgs
self-couplings were recently presented in ref.~\cite{Bahl:2025wzj},
but they were applied to a singlet extension of the SM. Also, a
two-loop calculation of the $h\rightarrow \gamma \gamma$ width
analogous to the one in ref.~\cite{Degrassi:2023eii} was performed in
ref.~\cite{Degrassi:2024qsf} for a triplet extension of the SM.}  In
all cases it was found that the two-loop corrections can significantly
modify the one-loop predictions of the THDM when the quartic couplings
involved are large, and should thus be taken into account for a
precise determination of the considered observable.

The trilinear Higgs coupling $\lambda_{hhh}$, which is crucial for
determining the shape of the Higgs potential and the dynamics of the
EW phase transition in the early universe, can be probed at the LHC
through the production of pairs of Higgs bosons. In the SM, the
amplitude for the main production process, $gg\rightarrow hh$,
comprises a direct ``box diagram'' contribution mediated by a quark
loop, plus a ``triangle diagram'' contribution with a quark-mediated
gluon-gluon-Higgs vertex, the $s$-channel exchange of a virtual scalar
$h$, and the production of an $hh$ pair through the trilinear coupling
$\lambda_{hhh}$. While the current sensitivity of the searches for
Higgs pair production at the LHC is still far from the SM prediction
for the corresponding cross section, those searches, also combined
with the measurements of single Higgs production, can provide
meaningful bounds on the deviations from the (tree-level) prediction
$\lambda^\smallSM_{hhh} = 3\,M_h^2/v$ ($v \approx 246$~GeV being the
Higgs vev) that might occur in BSM models. Indeed, the
ATLAS~\cite{ATLAS:2024ish} and CMS~\cite{CMS:2024awa} collaborations
already constrain the ratio $\kappa_\lambda \equiv
\lambda_{hhh}/\lambda_{hhh}^\smallSM$ to be in the ranges
$-1.2<\kappa_\lambda<7.2$ and $-1.2<\kappa_\lambda<7.5$, respectively,
under the assumption that $\lambda_{hhh}$ is the only coupling that
deviates from the SM prediction.

In the THDM, it has long been known (see, e.g.,
refs.~\cite{Kanemura:2002vm, Kanemura:2004mg}) that one-loop
corrections involving the BSM Higgs bosons can induce modifications to
$\lambda_{hhh}$ of ${\cal O}(100\%)$, or even larger. More recently,
ref.~\cite{Bahl:2022jnx} showed that there are regions of the THDM
parameter space -- still allowed by all theory-driven constraints on
the quartic Higgs couplings -- in which the two-loop corrections to
$\lambda_{hhh}$ computed in refs.~\cite{Braathen:2019pxr,
  Braathen:2019zoh} can be about as large as the one-loop
corrections. Thus, the inclusion of the two-loop corrections can
significantly alter the constraints on the THDM parameter space that
arise from the experimental bounds on $\kappa_\lambda$. Moreover, in
this model the $gg\rightarrow hh$ amplitude includes also a triangle
contribution involving the $s$-channel exchange of a heavy scalar $H$,
which can in turn have a significant impact on the prediction for the
pair-production cross section. While the coupling between two light
scalars and one heavy scalar, $\lambda_{hhH}$, vanishes at the
tree-level in the alignment limit of the THDM (i.e., when $h$ is
SM-like), a non-vanishing coupling is still generated by radiative
corrections. The impact of the one-loop corrections to the trilinear
couplings on the cross section for Higgs pair production at the LHC
was recently studied in ref.~\cite{Heinemeyer:2024hxa}, but the impact
of the corresponding two-loop corrections has not been addressed so
far.

\bigskip

In this paper, we study the impact of the two-loop corrections
controlled by the BSM Higgs couplings on the cross section for the
production of a pair of SM-like Higgs bosons via gluon fusion in the
aligned THDM. To this aim, we reassess the two-loop calculation of
$\lambda_{hhh}$, but we also consider additional contributions that,
in principle, affect the cross section at the same perturbative order
in the potentially-large couplings. In particular, we compute for the
first time the two-loop corrections to $\lambda_{hhH}$, and we include
also the relevant corrections to the Higgs-gluon couplings and to the
$s$-channel propagators. We provide explicit analytic formulas for the
two-loop corrections to the trilinear couplings in two simplified
scenarios, while making our general results available on request in
electronic form. We then discuss the numerical impact of the two-loop
BSM contributions, first on the individual couplings and then on the
prediction for the pair-production cross section. In particular, we
highlight the dependence of our results on the choice of
renormalization scheme and scale for one of the parameters that
determine the masses of the BSM Higgs bosons.

\vfill
\newpage

\section{Higgs pair production cross section in the aligned THDM}
\label{sec:setup}

In this section we summarize some general results on the production of
a pair of SM-like neutral scalars via gluon fusion in the aligned
THDM. Since we focus on the effect of the potentially large BSM Higgs
couplings, we only work at the leading order (LO) in QCD, and we also
neglect all contributions beyond the LO that are controlled by the
Yukawa couplings or by the EW gauge couplings. Therefore, in what
follows, our use of the terms LO, next-to-LO (NLO) and next-to-NLO
(NNLO) refers to an expansion in powers of the scalar couplings.  We
point the reader to section~2 of ref.~\cite{Degrassi:2023eii} for an
overview of the Higgs sector of the CP-conserving THDM, and for a
discussion of how the alignment condition must be imposed at the
perturbative order relevant to our calculations.
Additional details on the renormalization of the Higgs
sector that will be relevant to the two-loop calculation of the
trilinear Higgs couplings are provided in the appendix.

\bigskip

We can write the cross section for the production of an $hh$
pair via gluon fusion at the LHC as
\beq
M_{hh}^2\, \frac{d\sigma}{dM_{hh}^2}
~=~
\int_{\tau}^1\frac{dx}{x} \,f_g(x, \mu_\smallF)\, f_g(\tau/x, \mu_\smallF)\,
M_{hh}^2\,
\frac{d\hat{\sigma}}{dM_{hh}^2}~,
\eeq
where: $M_{hh}^2$ is the invariant squared mass of the $hh$ pair;
$f_g(x, \mu_\smallF)$ is the density of the gluon in the colliding
proton, $\mu_\smallF$ being the factorization scale; $\tau =
M_{hh}^2/s$, where $\sqrt s$ is the total center-of-mass energy;
$d\hat\sigma/dM_{hh}^2$ is the differential partonic cross section for
$gg\rightarrow hh$. For the latter, we have
\beq
\label{eq:partxs}
M_{hh}^2\,\frac{d\hat{\sigma}}{dM_{hh}^2}~=~
\frac{G_\mu^2\,\alpha_s^2(\mu_\smallR)}{512\, (2 \pi)^3} \int_{\hat{t}_-}^{\hat{t}_+} d\hat{t}\, \left(|\mathcal{F}|^2+|\mathcal{G}|^2\right)~,
\eeq
where: $G_\mu$ is the Fermi constant, with $v=(\sqrt 2 G_\mu)^{-1/2}$
at the tree level; $\alpha_s(\mu_\smallR)$ is the strong gauge
coupling expressed in the $\msbar$ renormalization scheme at the scale
$\mu_\smallR$; the Mandelstam variable $\hat t$ for the partonic
process is defined as
\beq
\hat t ~=~ M_h^2 - \frac{M_{hh}^2}{2}\,\left(1-\cos\theta\,
\sqrt{1-\frac{4M_h^2}{M_{hh}^2}}\,\right)~,
\label{eq:that}
\eeq
$\theta$ being the scattering angle in the partonic center-of-mass
system; the integration limits $\hat t_{\pm}$ are obtained by setting
$\cos\theta = \pm 1$ in eq.~(\ref{eq:that}) above; finally,
$\mathcal{F}$ and $\mathcal{G}$ represent the spin-zero and spin-two
form factors for the process $gg\rightarrow hh$, respectively.  While
the spin-two form factor ${\cal G}$ receives only contributions from
box diagrams, hence we write ${\cal G} = G_{\Box}$, the spin-zero form
factor ${\cal F}$ can be decomposed in box and triangle contributions
as
\beq
\label{eq:boxtriangle}
{\cal F} ~=~
F_{\Box}  ~+~C_{\Delta}^h\,F_{\Delta}^{h} ~+~  C_{\Delta}^H\,F_{\Delta}^{H}~.
\eeq
In particular, $F_{\Box}$ contains the spin-zero part of the box
diagrams, while $F_{\Delta}^{h}$ ($F_{\Delta}^{H}$) contains the
contribution of the triangle diagrams for the production of an
off-shell scalar $h$ ($H$) which subsequently turns into the $hh$
pair through the factor $C_{\Delta}^h$ ($C_{\Delta}^H$).

\paragraph{LO contributions:}
At the LO in the scalar couplings, the triangle and box form factors of
the THDM are related to the corresponding SM
quantities~\cite{Glover:1987nx,Plehn:1996wb} by a simple rescaling of
the Higgs--top--top coupling (we neglect here the small contributions
involving the lighter quarks):
\beq
F_\Delta^h ~=~ g_{htt}\,F_\Delta^\smallSM\,,~~~~
F_\Delta^H ~=~ g_{Htt}\,F_\Delta^\smallSM\,,~~~~
F_\Box ~=~ g_{htt}^2\,F_\Box^\smallSM\,,~~~~
G_\Box ~=~ g_{htt}^2\,G_\Box^\smallSM~.
\eeq
In the alignment limit, the rescaling factors become $g_{htt}=1$ and
$g_{Htt}=-\cot\beta$, where $\tan\beta \equiv v_2/v_1$ is defined as
the ratio of the vevs of the $SU(2)$ doublets in the basis $(\Phi_1,\Phi_2)$
in which a $Z_2$ symmetry forbids flavor-changing neutral-current
interactions at the tree level. Also at the LO in the scalar
couplings, the triangle form factors in eq.~(\ref{eq:boxtriangle}) are
multiplied by combinations of $s$-channel propagators and trilinear
couplings
\beq
\label{eq:Cfactors}
C_\Delta^h ~=~ \frac{\lambda_{hhh}^{\rm tree}\,v}{M_{hh}^2-M_h^2}~,
~~~~~~~~~~
C_\Delta^H ~=~ \frac{\lambda_{hhH}^{\rm tree}\,v}{M_{hh}^2-M_H^2 + i\,M_H\,\Gamma_H}~,
\eeq
where the width $\Gamma_H$ cures the pole in the $H$ propagator for
$M_{hh}^2\approx M_H^2$ (in contrast, there is no pole in the $h$
propagator because $M_{hh}^2 > 4M_h^2\,$). In the alignment limit we
have $\lambda_{hhh}^{\rm tree} = 3\,M_h^2/v$ and $\lambda_{hhH}^{\rm
  tree} = 0$, hence there is no LO contribution from the diagram with
$s$-channel exchange of $H$.

\bigskip

\paragraph{NLO contributions:}
Beyond the LO, we choose to focus on the corrections that involve the
highest powers of $\tilde\lambda$\,, by which we mean generic
combinations of quartic scalar couplings that can in principle be of
${\cal O}(1$--$10)$ without violating any constraints (the
corresponding trilinear couplings are then of the form
$\tilde\lambda\,v$). Now, both the one-loop corrections to the
$s$-channel propagators and the two-loop contributions to the box and
triangle form factors (for which the LO is at the one-loop level) are
at most of ${\cal O}(\tilde\lambda^2)$. In contrast, the trilinear
coupling $\lambda_{hhh}$ receives one-loop corrections of ${\cal
  O}(\tilde\lambda^3)$ which, as found in refs.~\cite{Kanemura:2002vm,
  Kanemura:2004mg}, can significantly exceed its tree-level
value. Similarly, $\lambda_{hhH}$ receives one-loop corrections of
${\cal O}(\tilde\lambda^3)$ which do not vanish even in the alignment
limit of the THDM. Therefore, the dominant NLO corrections controlled
by the potentially large BSM couplings are obtained by writing the
form factors ${\cal F}$ and ${\cal G}$ as in
eqs.~(\ref{eq:boxtriangle})--(\ref{eq:Cfactors}), with the sole
replacements $\lambda_{hhh}^{\rm tree} ~\longrightarrow
~\lambda_{hhh}^{1\ell}$\, and $\lambda_{hhH}^{\rm tree}
~\longrightarrow ~ \lambda_{hhH}^{1\ell}$\, in
eq.~(\ref{eq:Cfactors}). We will furthermore assume that
$\tilde\lambda \gg \lambda_{\smallSM}$, where $\lambda_{\smallSM} =
M_h^2/v^2 \approx 0.26$, and thus employ the approximation
$M_h^2\approx 0$ in all contributions beyond the LO.

The one-loop corrections to the trilinear couplings entering
eq.~(\ref{eq:Cfactors}) include genuine vertex corrections and
counterterm contributions stemming from the renormalization of the
tree-level couplings. To determine the latter, we remark that in the
alignment limit the tree-level couplings can be written as
$\lambda_{hhh}^{\rm tree}= 3 \Lambda_1 v$ and $\lambda_{hhH}^{\rm
  tree}= -3 \Lambda_6 v$, where $\Lambda_1$ and $\Lambda_6$ are
quartic couplings in the so-called Higgs basis, in which one doublet,
$\Phi_\smallSM$, develops the full SM-like vev $v$, while the other
doublet, $\Phi_\smallBSM$, has vanishing vev (see the appendix for
explicit definitions). At the tree level, the alignment condition
implies $\Lambda_1 = M_h^2/v^2$ and $\Lambda_6=0$.  However, as
discussed in ref.~\cite{Degrassi:2023eii}, we require that the
alignment condition apply to the radiatively corrected mass matrix for
the neutral scalars $h$ and $H$, computed at an external momentum
equal to the SM-like Higgs mass. Under the approximation $M_h^2\approx
0$, eqs.~(28) and (29) of ref.~\cite{Degrassi:2023eii} lead to
\bea
\label{eq:dlamhhh1l}
\delta^{1\ell}\lambda_{hhh} &=& \Lambda^{1\ell,\,\smallIPI}_{hhh}(M^2_{hh},0,0)
- \frac{3}{v}\left[\Pi^{1\ell}_{hh}(0) - \frac{T^{1\ell}_h}{v}\,\right]~,\\[2mm]
\label{eq:dlamhhH1l}
\delta^{1\ell}\lambda_{hhH} &=& \Lambda^{1\ell,\,\smallIPI}_{hhH}(M^2_{hh},0,0)
- \frac{3}{v}\left[\Pi^{1\ell}_{hH}(0) - \frac{T^{1\ell}_H}{v}\,\right]~,
\eea
where: we define $\lambda_{hh\phi}^{1\ell} = \lambda_{hh\phi}^{\rm
  tree}\, +\, \delta^{1\ell}\lambda_{hh\phi}$ for $\phi=(h,H)$;
$\Lambda^{1\ell,\,\smallIPI}_{hh\phi}(q^2,p_1^2,p_2^2)$ are the
one-loop one-particle irreducible (1PI) contributions to the $hh\phi$
vertex (note that $p_1$ and $p_2$ are the momenta on the external
legs, and $q=p_1+p_2$ is the momentum flowing in the $s$ channel);
$\Pi^{1\ell}_{h\phi}(p^2)$ and $T^{1\ell}_\phi$ are one-loop
self-energies and tadpoles, explicit expressions for which are given
in the appendix~A of ref.~\cite{Degrassi:2023eii}. The terms within
square brackets in eq.~(\ref{eq:dlamhhh1l}) stem from expressing
$\lambda_{hhh}^{\rm tree}$ in terms of the pole mass of the SM-like
Higgs boson, whereas the terms within square brackets in
eq.~(\ref{eq:dlamhhH1l}) stem from requiring that the alignment
condition apply beyond the tree level.  By direct computation of the
one-loop vertex diagrams under the appropriate limits, we obtain:
\bea
\label{eq:dlamhhh1lexpl}
\delta^{1\ell}\lambda_{hhh} &=& -\frac{1}{8\pi^2v^3}\biggr[
(\mHq-\MM)^2\,f_V(\mHq,\MM,M^2_{hh})\,+\,
  (\mAq-\MM)^2\,f_V(\mAq,\MM,M^2_{hh})\nn\\
&& ~~~~~~~~~~~~~ +
  2\,(\mHpq-\MM)^2\,f_V(\mHpq,\MM,M^2_{hh})\biggr]~,\\[2mm]
\label{eq:dlamhhH1lexpl}
\delta^{1\ell}\lambda_{hhH} &=& \frac{\cot2\beta}{8\pi^2v^3}\,(\mHq-\MM)\,
\biggr[
3\,(\mHq-\MM)\,f_V(\mHq,\MM,M^2_{hh})\,+\,
  (\mAq-\MM)\,f_V(\mAq,\MM,M^2_{hh})\nn\\
&& ~~~~~~~~~~~~~~~~~~~~~~~~~~~ +
  2\,(\mHpq-\MM)\,f_V(\mHpq,\MM,M^2_{hh})\biggr]~,
\eea
with
\beq
\label{eq:fV}
f_V(M_\Phi^2,\MM,M^2_{hh})\,\equiv\,4\,(M_\Phi^2-\MM)\,C_0(M^2_{hh},0,0;
M_\Phi^2,M_\Phi^2,M_\Phi^2)
+B_0(M^2_{hh};M_\Phi^2,M_\Phi^2)-B_0(0;M_\Phi^2,M_\Phi^2)\,,
\eeq
where $B_0(q^2;m_1^2,m_2^2)$ and
$C_0(q^2,p_1^2,p_2^2;m_1^2,m_2^2,m_3^2)$ are standard
Passarino-Veltman functions~\cite{Passarino:1978jh}. In
eqs.~(\ref{eq:dlamhhh1lexpl})--(\ref{eq:fV}), the minimum conditions
of the scalar potential, the alignment condition, and the
approximation $M_h^2\approx0$ have been used to write the couplings of
the BSM scalars $\Phi = (H,A,H^\pm)$ in terms of the mass differences
$(M_\Phi^2-M^2)$, where $M^2 \equiv 2\,m_{12}^2/\sin2\beta$, and
$m_{12}^2\,(\Phi_1^\dagger \Phi_2 + {\rm h.c.})$ is the mass term
that violates the $Z_2$ symmetry softly in the THDM Lagrangian.

\bigskip

We note that in eqs.~(\ref{eq:dlamhhh1l}) and (\ref{eq:dlamhhH1l})
there are no contributions involving the diagonal
wave-function-renormalization (WFR) counterterms $\delta Z_{hh}$ and
$\delta Z_{HH}$, because those counterterms would multiply the
tree-level Higgs couplings, but $\lambda_{hhh}^{\rm tree}$ vanishes
under the approximation $M_h^2\approx 0$ and $\lambda_{hhH}^{\rm
  tree}$ vanishes under the alignment condition. We also note that in
our derivation of eqs.~(\ref{eq:dlamhhh1l}) and (\ref{eq:dlamhhH1l})
we did not consider any contributions involving counterterms of mixing
angles, or the off-diagonal WFR counterterms $\delta Z_{Hh}$ and
$\delta Z_{hH}$. As discussed in ref.~\cite{Degrassi:2023eii}, since
our calculation is restricted to the alignment limit, we avoid
introducing a mixing angle for the neutral scalar sector, and just
require that the alignment still holds after the inclusion of
radiative corrections.  We will now show that, in the standard
approach to mixing renormalization, the WFR contributions combine with
the counterterms of the mixing angles in such a way that the final
result is the same as in our approach. Indeed, for generic mixing
between the neutral scalars, the tree-level trilinear couplings can be
written as
\bea
\label{eq:lhhhalpha}
\lambda_{hhh}^{\rm tree} &=&\!
-\frac{6}{v}\,\biggr[\,\frac{M_h^2}{2}\sin(\alpha-\beta) +
  \left(M_h^2-M^2\right)\,\cos^2(\alpha-\beta)
  \biggr(\sin(\alpha-\beta)-\cos(\alpha-\beta)\cot2\beta\biggr)\biggr],\\[2mm]
\label{eq:lhhHalpha}
\lambda_{hhH}^{\rm tree} &=&\!\frac{\cos(\alpha-\beta)}{v\,\sin2\beta}
\,\biggr[(2\,M_h^2+M_H^2-3\,M^2)\,\sin2\alpha + M^2\,\sin2\beta\biggr]~,
\eea
where $\alpha$ is the angle that rotates the neutral, CP-even
components of $\Phi_1$ and $\Phi_2$ into the mass eigenstates $H$ and
$h$, while the combination $(\alpha-\beta)$ rotates the neutral,
CP-even components of $\Phi_\smallSM$ and $\Phi_\smallBSM$ into $h$
and $H$. The alignment condition corresponds to $\alpha = \beta -
\pi/2$, in which case $\phi^0_\smallSM = h$ and $\phi^0_\smallBSM =
-H$.  It is easy to see from eqs.~(\ref{eq:lhhhalpha}) and
(\ref{eq:lhhHalpha}) that in this limit there are no contributions
from mixing renormalization to $\delta^{1\ell}\lambda_{hhh}$. In
contrast, the contributions to $\delta^{1\ell}\lambda_{hhH}$ become
\beq
\left[\, \frac{\partial \lambda_{hhH}^{\rm tree}}{\partial \alpha}\,\delta\alpha
  \,+\, \frac{\partial \lambda_{hhH}^{\rm tree}}{\partial \beta}\,\delta\beta
  \,+\, \lambda_{hHH}^{\rm tree}\,\delta Z_{Hh}\,
  \right]_{\alpha=\beta-\frac{\pi}2}
\!=~~ \frac{4\,M^2-M_H^2}{v}\,
\delta(\alpha-\beta)~-~
\frac{4(M^2-M_H^2)}{v}\,\frac{\delta Z_{Hh}}2~,
\label{eq:mixrenorm}
\eeq
where the approximation $M_h^2\approx0$ allowed us to neglect an
additional contribution $\lambda_{hhh}^{\rm tree}\,\delta
Z_{hH}/2$. Following ref.~\cite{Degrassi:2023eii}, we fix the
counterterm for $(\alpha-\beta)$ from the requirement that this
combination of angles diagonalize the loop-corrected mass matrix for
the neutral scalars in the Higgs basis, evaluated at the external
momentum $p^2=M_h^2$. In turn, the WFR counterterm $\delta Z_{Hh}$ is
fixed from the requirement that the renormalized two-point correlation
function $\Gamma_{hH}(p^2)$ vanish for $p^2=M_h^2$\,. In the alignment
limit, and taking $M_h^2=0$ for consistency with the rest of our
calculation, this leads to
\beq
\delta(\alpha-\beta)~=~\frac{\delta Z_{Hh}}2
~=~\frac{-1}{M_H^2}\,\left[\,
  \Pi^{1\ell}_{hH}(0)\,-\,\frac{T^{1\ell}_H}{v}\,\right]~,
\label{eq:mixrenorm2}
\eeq
hence the two terms on the r.h.s.~of eq.~(\ref{eq:mixrenorm}) combine
to reproduce the counterterm in eq.~(\ref{eq:dlamhhH1l}).

We remark that, at this stage, we only need to specify a renormalization
condition for the combination $(\alpha-\beta)$. However, even in the
alignment limit the one-loop correction to $\lambda_{hhH}$ depends
explicitly on $\beta$, see eq.~(\ref{eq:dlamhhH1lexpl}). In the
calculation of the two-loop correction to $\lambda_{hhH}$, it will
thus be necessary to specify a separate renormalization condition for
$\beta$. This implicitly defines a renormalization scheme for
$\alpha$, which may in principle differ from the schemes introduced,
e.g., in refs.~\cite{Kanemura:2004mg, Krause:2016oke,
  Altenkamp:2017ldc, Kanemura:2017wtm}, where the counterterms of the
mixing angles involve also self-energies computed at external momenta
equal to the heavy Higgs masses. Since in our two-loop calculation of
the trilinear couplings we will rely on the approximation of vanishing
external momenta, we find our approach, in which the mixing
counterterms only involve self-energies computed at vanishing momenta,
to be particularly advantageous.
Finally, we remark that the complications related to the gauge
dependence of scalar-mixing renormalization that were discussed in
refs.~\cite{Krause:2016oke, Altenkamp:2017ldc, Kanemura:2017wtm} do
not apply here, because all of our calculations are restricted for
simplicity to the limit of vanishing EW gauge couplings.

\paragraph{NNLO contributions:}
At the NNLO in the potentially large scalar couplings, the dominant
contributions to the triangle diagrams are of ${\cal
  O}(\tilde\lambda^5)$. Those can arise either directly, from the
two-loop corrections to the trilinear Higgs couplings, or indirectly,
as the product of one-loop, ${\cal O}(\tilde\lambda^3)$ corrections to
the trilinear couplings times ${\cal O}(\tilde\lambda^2)$ corrections
to other parts of the $gg\rightarrow hh$ amplitude -- namely, the
gluon--gluon--Higgs vertex and the $s$-channel propagators.
For what concerns the box diagrams, we must take into account that at
the LO their contribution to the $gg\rightarrow hh$ amplitude is
similar in size (and opposite in sign) to the contribution of the
triangle diagrams, and at the NLO they receive corrections of ${\cal
  O}(\tilde\lambda^2)$. Since we are considering scenarios in which
the NLO triangle diagrams are comparable with, if not larger than, the
LO triangle diagrams, we can expect the NLO box diagrams to be
comparable with, if not smaller than, the NNLO triangle diagrams.

In summary, for a determination of the Higgs pair-production cross
section at the NNLO in the scalar couplings we need to compute
two-loop corrections to the three-Higgs vertices and to the
gluon--gluon--Higgs (triangle) and gluon--gluon--Higgs--Higgs (box)
vertices, plus a number of contributions that stem from the product of
one-loop corrections. However, two-loop diagrams with three external
legs can be computed with relative ease only in the limit of vanishing
external momenta. To preserve the one-loop momentum dependence of the
trilinear Higgs couplings (i.e., their dependence on $M_{hh}^2$, since
we take the approximation $M_h^2\approx 0$ on the external $h$ legs),
while including the two-loop corrections computed at vanishing
momenta, we define the two-loop couplings $\lambda_{hh\phi}^{2\ell}$
for $\phi=(h,H)$ as follows:
\beq
\lambda_{hh\phi}^{2\ell} ~=~ \lambda_{hh\phi}^{\rm
  tree}\, +\, \delta^{1\ell}\lambda_{hh\phi}\,
\left( 1 + \left.\frac{\delta^{2\ell}\lambda_{hh\phi}}{\delta^{1\ell}\lambda_{hh\phi}}
\right|_{M_{hh}^2=0}\right)~,
\label{eq:lambda2l}
\eeq
where the calculation of $\delta^{2\ell}\lambda_{hh\phi}$ will be
described in sections~\ref{sec:lamhhh} and \ref{sec:lamhhH}, and
formulas for $\delta^{1\ell}\lambda_{hh\phi}$ in the limit
$M_{hh}^2=0$ can be obtained from
eqs.~(\ref{eq:dlamhhh1lexpl})--(\ref{eq:fV}) by considering that
$C_0(0,0,0;M_\Phi^2,M_\Phi^2,M_\Phi^2) = -1/(2\,M_\Phi^2)$. Similarly,
the triangle form factors at the NLO in the scalar couplings become,
in the alignment limit,
\beq
(F_\Delta^h)^\smallNLO ~=~ \,F_\Delta^\smallSM
\left(1 \,+\, \left.\frac{F_\Delta^{h,\,2\ell}}{F_\Delta^{h,\,1\ell}}\right|_{M_{hh}^2=0}
\right)\,,~~~~
(F_\Delta^H)^\smallNLO ~=~ -\cot\beta\,F_\Delta^\smallSM
\left(1 \,+\, \left.\frac{F_\Delta^{H,\,2\ell}}{F_\Delta^{H,\,1\ell}}\right|_{M_{hh}^2=0}
\right)~,
\label{eq:FDelNLO}
\eeq
where $F_\Delta^\smallSM$ retains the full momentum dependence, while
in the limit $M_{hh}^2=0$ we have $F_\Delta^{h,\,1\ell} = 4\,T_F/3$
and $F_\Delta^{H,\,1\ell} = -4\,T_F\,\cot\beta/3$, with $T_F=1/2$ a
color factor. The two-loop corrections to the spin-zero part of the
box diagrams, $F_\Box$, are in turn included for vanishing external
momenta
\beq
(F_\Box)^\smallNLO ~=~ \,F_\Box^\smallSM
\left(1 \,+\, \left.\frac{F_\Box^{2\ell}}{F_\Box^{1\ell}}\right|_{M_{hh}^2=0}
\right)~,
\label{eq:FBoxNLO}
\eeq
where again $F_\Box^\smallSM$ retains the full momentum dependence,
while for vanishing external momenta we have $F_\Box^{1\ell} =
-4\,T_F/3$.
The spin-two part of the box form factor vanishes under the
approximation of vanishing external momenta that we adopt in our
two-loop calculations.
The calculation of the ${\cal O}(\tilde\lambda^2)$ corrections to the
triangle and box form factors will be described in
section~\ref{sec:gluglu}.

The last source of ${\cal O}(\tilde\lambda^5)$ contributions to the
$gg\rightarrow hh$ amplitude are ${\cal O}(\tilde\lambda^2)$
corrections to the $s$-channel propagators combined with ${\cal
  O}(\tilde\lambda^3)$ corrections to the trilinear Higgs couplings.
At the NNLO in the scalar couplings, the factors $C_\Delta^h$ and
$C_\Delta^H$ in eq.~(\ref{eq:boxtriangle}) become
\bea
\label{eq:ChNNLO}
C_\Delta^h
&=& \frac{\lambda_{hhh}^{2\ell}\,v}{M_{hh}^2-M_h^2}\,
\left[ \,1 \,+\, \frac{\Pi^{1\ell}_{hh}(M_{hh}^2)-{\rm Re}\,\Pi^{1\ell}_{hh}(M_h^2)}{M_{hh}^2-M_h^2}\,\right]\nn\\[2mm]
&+ & \frac{\lambda_{hhH}^{1\ell}\,v}{M_{hh}^2-M_H^2 + i\,M_H\,\Gamma_H}\,
\left[ \,\frac{\Pi^{1\ell}_{hH}(M_{hh}^2)-\Pi^{1\ell}_{hH}(M_h^2)}{M_{hh}^2-M_h^2}\,\right]~,\\[5mm]
\label{eq:CHNNLO}
C_\Delta^H &=&
\frac{\lambda_{hhH}^{2\ell}\,v}{M_{hh}^2-M_H^2 + i\,M_H\,\Gamma_H}\,
\left[ \,1 \,+\, \frac{\Pi^{1\ell}_{HH}(M_{hh}^2)-\Pi^{1\ell}_{HH}(M_H^2)}{M_{hh}^2-M_H^2 + i\,M_H\,\Gamma_H}\,\right]\nn\\[2mm]
&+ & \frac{\lambda_{hhh}^{1\ell}\,v}{M_{hh}^2-M_H^2 + i\,M_H\,\Gamma_H}\,
\left[ \,\frac{\Pi^{1\ell}_{hH}(M_{hh}^2)-\Pi^{1\ell}_{hH}(M_h^2)}{M_{hh}^2-M_h^2}\,\right]~,
\eea
where, in accordance with our approximations, we can further set
$M_h^2 = 0$ in all of the terms within square brackets. While the
self-energies evaluated at $p^2=M_{hh}^2$ in eqs.~(\ref{eq:ChNNLO})
and (\ref{eq:CHNNLO}) stem from genuine loop insertions in the
$s$-channel propagators, the other self-energies stem from counterterm
contributions associated with the renormalization of the Higgs masses
and mixing. Note that, due to the presence of the width $\Gamma_H$ in
the NLO part of $C_\Delta^H$, the NNLO correction within square
brackets in the first line of eq.~(\ref{eq:CHNNLO}) includes the
whole $\Pi^{1\ell}_{HH}(M_H^2)$ rather than just the real
part. Finally, the form of the second lines of eqs.~(\ref{eq:ChNNLO})
and (\ref{eq:CHNNLO}) reflects the fact that, in our definition of the
alignment condition beyond the tree level, we require that the
$h$--$H$ mixing vanish for $p^2=M_h^2$, but the momentum flowing into
the off-diagonal self-energy is in fact $p^2=M_{hh}^2$.


\section{Two-loop corrections to the Higgs interactions}
\label{sec:twoloop}

In this section, we first reassess the calculation of the two-loop
corrections to $\lambda_{hhh}$ in the aligned THDM, which was provided
earlier in refs.~\cite{Braathen:2019pxr, Braathen:2019zoh}, then we
obtain new results for the two-loop corrections to $\lambda_{hhH}$,
also in the alignment limit. We provide explicit formulas for the
two-loop corrections to the trilinear couplings in two simplified
scenarios.  Finally, we consider the ${\cal O}(\tilde\lambda^2)$
corrections to the Higgs--gluon couplings, which stem from the
renormalization of $v$ and of the external Higgs legs.

Since we are focusing on the potentially large effects of the
corrections controlled by the quartic Higgs couplings, we neglect the
corrections controlled by the EW-gauge and Yukawa couplings, as they
are subdominant in the region of interest where some of the quartic
couplings are of ${\cal O}(1$--$10)$. We also treat the mass of the
SM-like Higgs boson as negligible w.r.t.~the masses of the BSM Higgs
bosons, and rely on the approximation of vanishing external momenta in
all two-loop diagrams.

\vfill
\newpage

\subsection{Two-loop corrections to $\lambda_{hhh}$}
\label{sec:lamhhh}

In the alignment limit of the THDM, the trilinear self-coupling of the
SM-like Higgs boson reduces to $\lambda^{\rm tree}_{hhh} = 3\,M_h^2/v$
at the tree level.  In the calculation of the two-loop corrections to
$\lambda_{hhh}$ we must specify a renormalization scheme for the
parameters entering both the tree-level and the one-loop parts of the
coupling. In particular, we assume that the tree-level coupling is
expressed in terms of the pole mass of the SM-like Higgs boson, and we
choose to express the one-loop part of the correction in terms of the
pole masses of the BSM Higgs bosons, the $\msbar$-renormalized
parameter $\MM$, and the Fermi constant $G_\mu$. Taking the limit
$M_{hh}^2 = 0$ in eq.~(\ref{eq:dlamhhh1lexpl}), we obtain the one-loop
correction at vanishing external momenta
\beq
\delta_0^{1\ell}\lambda_{hhh}~=~
\frac{(\sqrt2\,G_\mu)^{\frac32}}{4\pi^2}\,
\left[\,M_H^4\left(1-\frac{\MM}{M_H^2}\right)^3
  +\,M_A^4\left(1-\frac{\MM}{M_A^2}\right)^3
  +\,2\,M_{H^\pm}^4\left(1-\frac{\MM}{M_{H^\pm}^2}\right)^3\,\right]~,
\label{eq:lamhhh1lzeromom}
\eeq
where we henceforth denote $\delta_0 \Lambda \,\equiv\, \left.\delta
\Lambda\,\right|_{M_{hh}^2=0}$ for a generic momentum-dependent correction
$\delta\Lambda\,$.

Following the effective-potential approach outlined in
refs.~\cite{Braathen:2019pxr, Braathen:2019zoh}, the two-loop
correction to $\lambda_{hhh}$ can be computed from the derivatives of
the BSM-Higgs contribution to the two-loop effective potential, which
we denote as $\Delta V^{2\ell}$, w.r.t.~the lightest scalar field $h$.
In the alignment limit, where $h$ enters the effective potential only
in the combination $(v+h)$, the derivatives can be taken directly
w.r.t.~$v$, and we can also exploit the fact that the tree-level
squared masses of the BSM Higgs bosons $\Phi = (H,A,H^\pm)$ are all of
the form $M_\Phi^2 = \MM \,+\, \tilde\lambda_\phi\,v^2$, where $\tilde
\lambda_\phi$ are different combinations of quartic couplings. The
two-loop correction can thus be written as
\bea
\delta_0^{2\ell}\lambda_{hhh} &=&
 \left[ \frac{d^3}{dv^3}
  - \frac{3}{v}\left(\frac{d^2}{dv^2} - \frac1v\frac{d}{dv}\right)\right]
~\Delta V^{2\ell}\nn\\[3mm]
 &&\!\!- \!\!\sum_{{\scriptscriptstyle \Phi=(H,A,H^\pm)}}
 \!\!\frac{\partial (\delta_0^{1\ell}\lambda_{hhh})}
  {\partial M^2_\Phi}\, \delta M^2_\Phi
  ~+~ \frac{\partial (\delta_0^{1\ell}\lambda_{hhh})}
  {\partial \MM}\, (\delta \MM)_{\rm{\scriptscriptstyle cpl}}
  \,+\,\left(\delta Z_{hh} -
3\,\frac{\delta v}{v}\right)\,
  \delta_0^{1\ell}\lambda_{hhh}~,
\label{eq:dlam-2loop}
\eea
where the term in the first line that involves the third derivative of
$\Delta V^{2\ell}$ corresponds to the proper two-loop vertex
correction, while the terms that involve the second and first
derivatives are counterterm contributions -- analogous to the
self-energy and tadpole terms in eq.~(\ref{eq:dlamhhh1l}) -- stemming
from the requirement that $\lambda^{\rm tree}_{hhh}$ be expressed in
terms of the pole mass of the SM-like Higgs boson. We computed $\Delta
V^{2\ell}$ by adapting to the aligned THDM the formulas for the scalar
potential of a generic renormalizable theory provided in
ref.~\cite{Martin:2001vx}, and we obtained relatively compact
expressions for the derivatives of the two-loop integrals by means of
the recursive relation given in the appendix~A of
ref.~\cite{Dedes:2002dy}. We verified that our result for the first
line of eq.~(\ref{eq:dlam-2loop}) reproduces the one obtained by
setting $m_t=0$ in eq.~(B.7)\footnote{There appears to be a typo in
the 25th line of that equation, where $m_H^2$ should be replaced by
$m_A^2$.} of ref.~\cite{Braathen:2019zoh}, and we also verified that
the same result can be obtained through a direct calculation of the
relevant two-loop vertex, self-energy, and tadpole diagrams.

\newpage

The second line of eq.~(\ref{eq:dlam-2loop}) contains additional
counterterm contributions. In particular, the first term stems from
our scheme choice for the masses of the BSM Higgs bosons
$\Phi=(H,A,H^\pm)$ entering the one-loop part of the correction, and
explicit formulas for the mass counterterms $\delta M^2_\Phi =
{\rm Re}\,\Pi_{\Phi\Phi}(M_\Phi^2)$ are given in ref.~\cite{Braathen:2019zoh}.
The second term stems from the one-loop corrections to the relations
between the $h\Phi\Phi$ couplings entering the one-loop part of
$\lambda_{hhh}$, the BSM-Higgs masses $M_\Phi^2$, and the $\msbar$ parameter
$M^2$. The shift $(\delta \MM)_{\rm{\scriptscriptstyle cpl}}$, derived
earlier in ref.~\cite{Degrassi:2023eii}, can be written as
\beq
(\delta \MM)_{\rm{\scriptscriptstyle cpl}}
~=~
\frac12\,\left(\Pi_{hh}(0) - \frac{T^{1\ell}_h}{v}\right)
\,-\, \cot2\beta \,\left( \Pi^{1\ell}_{hH}(0) - \frac{T^{1\ell}_H}{v}\right)
\,-\,\frac{1}{v}\biggr(T^{1\ell}_h - 2\,\cot2\beta\,T^{1\ell}_H\biggr)~,
\label{eq:M22toM2}
\eeq
where the first term on the r.h.s.~stems from the renormalization of
$M_h^2$, the second term stems from the renormalization of
$\Lambda_6$, and the third term stems from the corrections to the
minimum conditions of the scalar potential
(see also the appendix).
Finally, $\delta Z_{hh}$ is the diagonal WFR counterterm for the two
external $h$ legs in the $gg\rightarrow hh$ amplitude, and $\delta v$
is the correction to the relation between $v$ and $G_\mu$, both
computed in the alignment limit and for $M_h^2\approx
0$~\cite{Degrassi:2023eii}:
\bea
\label{eq:dZhh}
\delta Z_{hh} &=& \left.\frac{d \Pi^{1\ell}_{hh}(p^2)}{dp^2}\right|_{p^2=0} \!=
-\frac{1}{48\pi^2v^2}\,\biggr[\,
  \frac{(\mHq-\MM)^2}{\mHq}~+~\frac{(\mAq-\MM)^2}{\mAq}
  ~+~2\,\frac{(\mHpq-\MM)^2}{\mHpq} \,\biggr],\\[3mm]
\label{eq:dv}
\frac{\delta v}{v} &=& ~~~\frac{\Pi^{1\ell}_{WW}(0)}{2\,M_W^2}~~~\,=~
-\frac{1}{8\pi^2v^2}\,\biggr[\,\wt{B}_{22}(0,\mHq,\mHpq)
  ~+~\wt{B}_{22}(0,\mAq,\mHpq)\,\biggr]~,
\eea
with
\beq
\widetilde B_{22}(0,m_1^2,m_2^2)~=~
\frac12\,\left(\frac{m_1^2+m_2^2}{4}-\frac{m_1^2\,m_2^2}{2\,(m_1^2-m_2^2)}
\,\ln\frac{m_1^2}{m_2^2}\right)~.
\eeq

Comparing our calculation of the two-loop corrections to
$\lambda_{hhh}$ with the one described in
ref.~\cite{Braathen:2019zoh}, we note that the latter omits the
contributions that arise from the one-loop corrections to the
relations between the BSM Higgs couplings and $\MM$, see
eq.~(\ref{eq:M22toM2}).\footnote{Other discrepancies between our
results for $\delta_0^{2\ell}\lambda_{hhh}$ and those in
ref.~\cite{Braathen:2019zoh} stem from the omission in eq.~(III.19) of
that paper of the BSM-Higgs contributions to $\delta v$, which vanish
only for degenerate BSM-Higgs masses, and from an error in the WFR
contribution in eq.~(V.14), where the last term should be divided by
$3$ instead of $2$.}

\bigskip

At this stage, it might again be worth commenting on the absence of
any contributions involving counterterms of mixing angles, or the
off-diagonal WFR counterterm $\delta Z_{Hh}$, from
eq.~(\ref{eq:dlam-2loop}). In our approach, the contributions to
$\delta_0^{2\ell}\lambda_{hhh}$ from diagrams with an off-diagonal
self-energy on an external leg cancel against counterterm insertions,
and the contributions that arise from the renormalization of
$\Lambda_6$ are all accounted for by terms included in $(\delta
\MM)_{\rm{\scriptscriptstyle cpl}}$, see eq.~(\ref{eq:M22toM2}). We
verified that, in the limit of vanishing Higgs mass and mixing, these
corrections coincide with the ones that we would obtain in the
standard approach to scalar-mixing renormalization, i.e., by combining
the contributions of mixing-angle counterterms and off-diagonal WFR
along the lines of
eqs.~(\ref{eq:lhhhalpha})--(\ref{eq:mixrenorm2}). To obtain the
necessary one-loop tadpoles, self-energies, and vertex corrections for
generic values of $\alpha$ and $\beta$, we adapted to the THDM the
general formulas provided in refs.~\cite{Braathen:2016cqe,
  Braathen:2018htl}.

\subsection{Two-loop corrections to $\lambda_{hhH}$}
\label{sec:lamhhH}

The calculation of the two-loop corrections to $\lambda_{hhH}$ entails
several complications when compared with the analogous calculation for
$\lambda_{hhh}$. First of all, in the alignment limit it would not be
possible to trade the derivative w.r.t.~the field $H$ with a
derivative w.r.t.~a vev, because the BSM Higgs doublet has no vev. To
circumvent this limitation, we compute the two-loop tadpole
$T_H^{2\ell}$ by adapting to the aligned THDM the results provided in
ref.~\cite{Braathen:2016cqe} for a general renormalizable theory, then
we take only two derivatives w.r.t.~$v$ to account for the remaining
$h$ legs. As in the case of the two-loop corrections to
$\lambda_{hhh}$, we verified that the same result can be obtained
through a direct calculation of the relevant two-loop vertex,
self-energy, and tadpole diagrams.  Second, while the trilinear
couplings entering the one-loop vertex correction $\delta_0^{1\ell}
\lambda_{hhh}$ in eq.~(\ref{eq:lamhhh1lzeromom}) are all of the
$h\Phi\Phi$ kind, with $\Phi=(H,A,H^\pm)$, the diagrams contributing
to the correction $\delta_0^{1\ell} \lambda_{hhH}$ involve two
$h\Phi\Phi$ couplings and one $H\Phi\Phi$ coupling. We thus need to
include a separate counterterm contribution for the relation between
the $H\Phi\Phi$ couplings, the mass $M_H^2$, and the $\msbar$
parameter $M^2$. Finally, by direct calculation of the counterterm
contributions associated to the renormalization of the mixing between
light and heavy fields, we find that, differently from the case of
$\lambda_{hhh}$, they are not all accounted for by the counterterms of
the $h\Phi\Phi$ and $H\Phi\Phi$ couplings. Indeed, we find two
additional mixing-induced counterterm contributions that need to be
explicitly included in our result for $\delta_0^{2\ell}\lambda_{hhH}$.

\bigskip

Taking the limit $M_{hh}^2 = 0$ in eq.~(\ref{eq:dlamhhH1lexpl}), we
obtain the one-loop correction to $\lambda_{hhH}$ at vanishing
external momenta
\bea
\delta_0^{1\ell}\lambda_{hhH}&=&
-\frac{(\sqrt2\,G_\mu)^{\frac32}}{4\pi^2}\,\cot2\beta\,(M_H^2 - \tilde M^2)
\left[3\,M_H^2\left(1-\frac{\MM}{M_H^2}\right)^2
  +\,M_A^2\left(1-\frac{\MM}{M_A^2}\right)^2\right.\nn\\[2mm]
  &&~~~~~~~~~~~~~~~~~~~~~~~~~~~~~~~~~~~~~~~~~~~~~~~~~~~~~~~~~~~~~~~~
  \left.+\,2\,M_{H^\pm}^2\left(1-\frac{\MM}{M_{H^\pm}^2}\right)^2\,\right]~,
\label{eq:lamhhH1lzeromom}
\eea
where the overall multiplicative factor $\cot2\beta\,(M_H^2 - \tilde
M^2)$ stems from the $H\Phi\Phi$ couplings, which in the alignment
limit are all controlled by the quartic coupling $\Lambda_7$ (see the
appendix). In analogy with the calculation of $\lambda_{hhh}$ in the
previous section, we express the one-loop corrections to the relation
between the trilinear couplings, the BSM Higgs masses, and $\MM$ as
shifts to the latter parameter.  However this requires that, in the
calculation of the counterterm contributions to
$\delta_0^{2\ell}\lambda_{hhH}$, we distinguish the parameter entering
the $H\Phi\Phi$ couplings, $\tilde M^2$, from the parameter entering
the $h\Phi\Phi$ couplings, $\MM$, as the two kinds of couplings
receive different shifts.  We also need to specify a renormalization
prescription for the angle $\beta$ that enters
$\delta_0^{1\ell}\lambda_{hhH}$ in eq.~(\ref{eq:lamhhH1lzeromom}) via
the $H\Phi\Phi$ couplings, and we choose to define it as an
$\msbar$-renormalized parameter. However, we remark that in the
approximation of vanishing Yukawa couplings there are no divergent
contributions to the counterterm for $\beta$, hence we are effectively
taking $\delta\beta=0$. In the standard approach to mixing
renormalization, this would also fix the counterterm $\delta\alpha$
when combined with our definition for the combination $(\alpha-\beta)$
in eq.~(\ref{eq:mixrenorm2}).

\vfill
\newpage

The two-loop correction to $\lambda_{hhH}$ can in turn be written as
\bea
\delta_0^{2\ell}\lambda_{hhH} &=&
 \left[ \frac{d^2}{dv^2}
  - \frac{3}{v}\left(\frac{d}{dv} - \frac1v\right)\right]
~T_H^{2\ell}\nn\\[3mm]
 && \!\!-\!\!\sum_{{\scriptscriptstyle \Phi=(H,A,H^\pm)}}
 \!\!\frac{\partial (\delta_0^{1\ell}\lambda_{hhH})}
  {\partial M^2_\Phi}\, \delta M^2_\Phi
  ~+~ \frac{\partial (\delta_0^{1\ell}\lambda_{hhH})}
  {\partial \MM}\, (\delta \MM)_{\rm{\scriptscriptstyle cpl}}
  ~+~ \frac{\partial (\delta_0^{1\ell}\lambda_{hhH})}
  {\partial \tilde M^2}\, (\delta \tilde M^2)_{\rm{\scriptscriptstyle cpl}}\nn\\[3mm]
  &+&\left(\delta Z_{hh} - 3\,\frac{\delta v}{v}\right)\,
  \delta_0^{1\ell}\lambda_{hhH} ~+~ \Delta_{\rm mix}\,\lambda_{hhH}~,
\label{eq:dlamhhH-2loop}
\eea
where the term in the first line that involves the second derivative
of $T_H^{2\ell}$ corresponds to the proper two-loop vertex correction,
while the other two terms are counterterm contributions -- analogous
to the self-energy and tadpole terms in eq.~(\ref{eq:dlamhhH1l}) --
stemming from the requirement that the alignment condition apply
beyond the tree level.

The first two terms in the second line of eq.~(\ref{eq:dlamhhH-2loop})
are analogous to those in eq.~(\ref{eq:dlam-2loop}), with $\delta
M^2_\Phi$ given in ref.~\cite{Braathen:2019zoh} and $(\delta
\MM)_{\rm{\scriptscriptstyle cpl}}$ given in
eq.~(\ref{eq:M22toM2}). The third term stems instead from the one-loop
corrections to the relations between the $H\Phi\Phi$ couplings
entering the one-loop part of $\lambda_{hhH}$, the mass $M_H^2$, and
the $\msbar$ parameter $M^2$. The shift $(\delta \tilde M^2)
_{\rm{\scriptscriptstyle cpl}}$
, whose derivation is described in the appendix,
reads
\beq
(\delta \tilde M^2)_{\rm{\scriptscriptstyle cpl}}
~=~ \,-\, \frac12\,\tan 2\beta
\,\left( \Pi^{1\ell}_{hH}(0) - \frac{T^{1\ell}_H}{v}\right)
\,-\,\frac{1}{v}\biggr(T^{1\ell}_h - 2\,\cot2\beta\,T^{1\ell}_H\biggr)~.
\label{eq:MtildetoM2}
\eeq
Note that after the inclusion of this contribution in
eq.~(\ref{eq:dlamhhH-2loop}) we can identify $\tilde M^2$ with $\MM$.

The terms that multiply $\delta_0^{1\ell}\lambda_{hhH}$ in the third
line of eq.~(\ref{eq:dlamhhH-2loop}) are analogous to those
multiplying $\delta_0^{1\ell}\lambda_{hhh}$ in the second line of
eq.~(\ref{eq:dlam-2loop}), with $\delta Z_{hh}$ and $\delta v$ defined
in eqs.~(\ref{eq:dZhh}) and (\ref{eq:dv}), respectively. The last term
in the third line of eq.~(\ref{eq:dlamhhH-2loop}) contains two
counterterm contributions associated with the renormalization of the
mixing between light and heavy scalars that are not accounted for by
the mixing-induced terms in the shifts $(\delta
M^2)_{\rm{\scriptscriptstyle cpl}}$ and $(\delta \tilde
M^2)_{\rm{\scriptscriptstyle cpl}}$\,. It reads

\bea
\Delta_{\rm mix}\,\lambda_{hhH} & = &
-~\frac{1}{4\pi^2v^3} \left[\,\frac{(\mAq-\mHq)(\mAq-\MM)^2}{M_A^4}\,\Pi^{1\ell}_{G^0A}(0)
  \right.\nn\\[2mm]
  &&~~~~~~~~~~~~~
  \left.+~\frac{2(\mHpq-\mHq)(\mHpq-\MM)^2}{M_{H^{\pm}}^4}\,\Pi^{1\ell}_{G^{\pm}H^{\pm}}(0)
  -~\frac{2(\mHq-\MM)^3}{M_H^4}\,\,\Pi^{1\ell}_{hH}(0)\,\right]\nn\\[3mm]
&&+~\frac{1}{4\pi^2v^3}\left[\,\frac{(\mAq-\mHq)(\mAq-\MM)}\mAq
  +\frac{2(\mHpq-\mHq)(\mHpq-\MM)}\mHpq\right.\nn\\[2mm]
  &&~~~~~~~~~~~~~~~~~~~~~~~~~~~~~~~~~~~~~~~~~~~~~~~~~~~~~~~
  \left.-~\frac{6(\mHq-\MM)^2}{M_H^2}\,\right]
\left(\Pi^{1\ell}_{hH}(0) - \frac{T^{1\ell}_H}{v}\right)~.\nn\\
\label{eq:dmix}
\eea

\vfill
\newpage

As detailed in the appendix, the terms in the first and second lines
of eq.~(\ref{eq:dmix}) stem from counterterm insertions that mix light
and heavy scalars in the internal lines of the loop diagrams. We
remark that $\Pi^{1\ell}_{G^0A}(0)= \Pi^{1\ell}_{G^{\pm}H^{\pm}}(0) =
T^{1\ell}_H/v$.
Finally, the third and fourth lines of eq.~(\ref{eq:dmix}) collect all
terms stemming from the renormalization of $\Lambda_6$ that are not
accounted for by $(\delta M^2)_{\rm{\scriptscriptstyle cpl}}$ and
$(\delta \tilde M^2) _{\rm{\scriptscriptstyle cpl}}$\,.

\vspace*{1mm}
\subsection{Explicit formulas for the trilinear couplings in two
  simplified scenarios} 
\label{sec:formulas}
\vspace*{1mm}

The formulas for the two-loop BSM-Higgs contributions to
$\lambda_{hhh}$ and $\lambda_{hhH}$, valid for generic values of all
the relevant masses, are rather lengthy and not particularly
illuminating. We thus make them available on request in electronic
form, and print explicit formulas only under two different sets of
simplifying assumptions. In the first case, we take the pole masses of
the BSM Higgs bosons to be degenerate, i.e., $M_H^2 = M_A^2 =
M_{H^\pm}^2 \equiv M_\Phi^2$. We thus obtain

\bea
\label{eq:2loophhh-deg}
\delta_0^{1\ell+2\ell}\lambda_{hhh} &=&
\frac{(\sqrt2\,G_\mu)^{\frac32}\,M_\Phi^4}{\pi^2}\,
\left(1-\frac{\MM}{M_\Phi^2}\right)^3
\nn\\[2mm]
&+& \frac{M_\Phi^6}{4\pi^4 v^5}\,
\left(1-\frac{\MM}{M_\Phi^2}\right)^3\,\biggr\{\,
\frac{9}{4} \,\cot^22\beta\, \left(1-\frac{\MM}{M_\Phi^2}\right)
\,\left[3 - \frac{\pi}{\sqrt 3}
\left(2+ \frac{\MM}{M_\Phi^2} \right)\right]\nn\\[2mm]
  && ~~~~~~~~~~~~~~~~~~~~~~~~~~~~
-\frac13 \left(1-\frac{\MM}{M_\Phi^2}\right)^2
-\frac{3\,\MM}{M_\Phi^2}\,
\left(1-\ln\frac{M_\Phi^2}{Q^2}\right)\,\biggr\}~,\\[5mm]
\label{eq:2loophhH-deg}
\delta_0^{1\ell+2\ell}\lambda_{hhH} &=&
-\frac{3(\sqrt2\,G_\mu)^{\frac32}\,M_\Phi^4}{2\,\pi^2}\,
\left(1-\frac{\MM}{M_\Phi^2}\right)^3\!\cot2\beta\nn\\[2mm]
&+& \frac{M_\Phi^6}{16\pi^4 v^5}
\left(1-\frac{\MM}{M_\Phi^2}\right)^3\!\cot2\beta\,\biggr\{\,
3\left(1-\frac{\MM}{M_\Phi^2}\right)\left[
  \frac{\pi}{\sqrt 3}\,\left(19 + 7\,\frac{\MM}{M_\Phi^2}\right)\,-\,32\right]
\cot^22\beta\nn\\[2mm]
&&~~~~~~~~~~~~~~~~~~~~~~~~~~~~~~~~~~~
+\left(1-\frac{\MM}{M_\Phi^2}\right)\!\left(11+16\,\frac{\MM}{M_\Phi^2}\right)
+\frac{18\,\MM}{M_\Phi^2}\left(1-\ln\frac{M_\Phi^2}{Q^2}\right)\biggr\}~,\nn\\
\eea
where the first line of each equation represents the one-loop part of
the correction, and the second and third lines represent the two-loop
part. In the last term within curly brackets of each equation, $Q^2$
represents the renormalization scale at which the $\msbar$ parameter
$\MM$ entering the one-loop part of the correction is expressed. For
later convenience, we also define a correction
$\delta_0^{1\ell+2\ell}\hat\lambda_{hhh}$ in which all three of the
$h$ legs are renormalized in the OS scheme. This can be obtained by
changing the pre-factor of the first term in the third line of
eq.~(\ref{eq:2loophhh-deg}) from $-1/3$ to $-1/2$.

A second limiting case, corresponding to the benchmark scenario
proposed in ref.~\cite{Bahl:2022jnx}, consists in taking $M_H^2 = \MM$
and the remaining two BSM-Higgs masses degenerate, i.e., $M_A^2 =
M_{H^\pm}^2 \equiv M_\Phi^2$. We thus obtain
\bea
\label{eq:2loophhh-ben}
\delta_0^{1\ell+2\ell}\lambda_{hhh} &=&
\frac{3\,(\sqrt2\,G_\mu)^{\frac32}\,M_\Phi^4}{4\pi^2}\,
\left(1-\frac{\MM}{M_\Phi^2}\right)^3
\nn\\[2mm]
&-& \frac{3\,M_\Phi^6}{256\pi^4 v^5}\,
\left(1-\frac{\MM}{M_\Phi^2}\right)^3\,\biggr\{\,
\frac{8\,M^4}{M_\Phi^4} + \frac{9\,\MM}{M_\Phi^2} - 23
+ 2\,\left(4+\frac{3\,\MM}{\MM-M_\Phi^2}\right)\,\ln\frac{\MM}{M_\Phi^2}
  \nn\\[2mm]
  && ~~~~~~~~~~~~~~~~~~~~~~~~~~~~~~\,
  +\,4\left(1-\frac{\MM}{M_\Phi^2}\right)^2\left(2+\frac{\MM}{M_\Phi^2}\right)
  \ln\left|1-\frac{M_\Phi^2}{\MM}\right|\nn\\[2mm]
  && ~~~~~~~~~~~~~~~~~~~~~~~~~~~~~~
 \,+\,\frac{36\,\MM}{M_\Phi^2}
 \left(1-\ln\frac{M_\Phi^2}{Q^2}\right)\,\biggr\}~,\\[5mm]
\label{eq:2loophhH-ben}
 \delta_0^{1\ell+2\ell}\lambda_{hhH} &=&
-\frac{9\,M_\Phi^6}{64\pi^4 v^5}
\left(1-\frac{\MM}{M_\Phi^2}\right)^3\!\cot2\beta\,\biggr\{\,
1 \,-\, \frac{\MM}{M_\Phi^2}\left[1-\left(1-\frac{M_\Phi^2}{\MM}\right)^2
  \ln\left|1-\frac{\MM}{M_\Phi^2}\right|\right]\nn\\[2mm]
  &&  ~~~~~~~~~~~~~~~~~~~~~~~~~~~~~~~~~~~~~~~
  -  \frac{\MM}{M_\Phi^2}\left(1-\ln\frac{M_\Phi^2}{Q^2}\right)\,\biggr\}~.
\eea

In this case $\delta_0^{1\ell+2\ell}\hat \lambda_{hhh}$ can be obtained by
changing the pre-factors of the first three terms within curly
brackets in the second line of eq.~(\ref{eq:2loophhh-ben}) from
$(8,9,-23)$ to $(10,5,-21)$.
We also remark that the correction to $\lambda_{hhh}$ in
eq.~(\ref{eq:2loophhh-ben}) does not depend on $\beta$, and that the
correction to $\lambda_{hhH}$ in eq.~(\ref{eq:2loophhH-ben}) starts
directly at the two-loop level. This is due to the fact that, in the
alignment limit, the tree-level couplings involving three heavy
scalars are all proportional to $(\mHq-\MM)\,\cot2\beta$, hence they
vanish in this simplified scenario. The non-vanishing two-loop
contribution to $\lambda_{hhH}$ stems from the fact that the condition
$\mHq=\MM$ is imposed on the pole mass of $H$ rather than on the
$\msbar$ mass, and from the presence in $(\delta \tilde
M^2)_{\rm{\scriptscriptstyle cpl}}$\,, see eq.~(\ref{eq:MtildetoM2}), of
terms that do not vanish for $\mHq=\MM$.

\bigskip

To conclude this section, we comment on the decoupling behavior of the
corrections to the trilinear couplings for large values of the mass
parameter $\MM$.  As discussed in refs.~\cite{Braathen:2019pxr,
  Braathen:2019zoh}, when $\MM$ is defined as an $\msbar$-renormalized
parameter the two-loop parts of the corrections to the trilinear
couplings do not vanish in the limit $\MM\rightarrow\infty$, when all
of the BSM-Higgs masses become very large. However, the sum of one-
and two-loop parts does vanish if the one-loop relations between $\MM$
and the Higgs masses are taken into account. To make the decoupling
behavior more explicit, the non-decoupling terms can be absorbed in a
redefinition of the mass parameter, $(\MM)^{\rm dec} \,=\,
(\MM)^{\msbar} \,+\, \delta \MM$, with~\cite{Degrassi:2023eii}
\beq
\label{eq:deltaM2}
\delta \MM ~=~ \frac{\MM}{16\pi^2\,v^2}\,\left[
  \left(\mHq+\mAq+2\,\mHpq-4\,\MM\right)
  \left(1-\ln\frac{\MM}{Q^2}\right)\,\right]~.
\eeq
If the one-loop parts of the corrections are expressed in terms of the
``decoupling'' mass parameter $(\MM)^{\rm dec}$, all terms of the form
$\left(1-\ln(M_\Phi^2/Q^2)\right)$ in
eqs.~(\ref{eq:2loophhh-deg})--(\ref{eq:2loophhH-ben}) are replaced by
$\ln(\MM/M_\Phi^2)\,$. In that case the two-loop parts of the
corrections do not depend explicitly on the renormalization scale, and
vanish when $\MM\rightarrow\infty$.

\subsection{Two-loop corrections to the Higgs--gluon interactions}
\label{sec:gluglu}

Among the two-loop contributions to the Higgs--gluon form factors
$F_\Delta^h$, $F_\Delta^H$, and $F_\Box$ introduced in
eq.~(\ref{eq:boxtriangle}), those that stem from 1PI diagrams
involving Higgs bosons coupled to the top-quark loop are ultimately of
${\cal O}(y_t^2)$, where $y_t$ is the top Yukawa coupling, and thus
cannot contribute to the cross section for Higgs-pair production at
the NNLO in the potentially large scalar couplings.
To verify this statement, we obtained
$F_\Delta^{h,\,2\ell,\,{\scriptscriptstyle {\rm 1PI}}}$ and
$F_\Box^{2\ell,\,{\scriptscriptstyle {\rm 1PI}}}$ at vanishing
external momenta from a
calculation along the lines of those performed for the MSSM in
refs.~\cite{Degrassi:2008zj, Agostini:2016vze}, relying on the
low-energy theorem for the Higgs--gluon interactions from
refs.~\cite{Shifman:1979eb, Kniehl:1995tn}.  In the case of
$F_\Delta^{H,\,2\ell,\,{\scriptscriptstyle {\rm 1PI}}}$ we computed
directly the two-loop 1PI contributions to the $ggH$ vertex at
vanishing external momentum.\footnote{We refrain from providing
details on these calculations since their results are ultimately not
included in our analysis.} To identify the terms that may contribute
at the highest order in the scalar couplings, we further expanded our
results for the triangle and box form factors in powers of
$m_t^2/M_\Phi^2$, with $\Phi = (H,A,H^\pm)$. We thus found:
  
  \bea
  \label{eq:FDh2l}
  F_\Delta^{h,\,2\ell,\,{\scriptscriptstyle {\rm 1PI}}} &=&
  \frac{m_t^2\,\cot^2\beta}{48\,\pi^2\,v^2}\,\left[\,
    3\left(1-\frac{M^2}{M_H^2}\right)
    -5\left(1-\frac{M^2}{M_A^2}\right)
    -4\left(1-\frac{M^2}{M_{H^\pm}^2}\right)
    \,+\,{\cal O}\left(\frac{m_t^2}{M_\Phi^2}\right)\,\right]~,\\[5mm]
  F_\Delta^{H,\,2\ell,\,{\scriptscriptstyle {\rm 1PI}}} &=&
  \label{eq:FDH2l}
   \frac{m_t^2\,\cot^2\beta}{48\,\pi^2\,v^2}\,\left\{\,
   \cot 2\beta\,\left(1-\frac{M^2}{M_H^2}\right)\left(- 9 +
   5\,\frac{M_H^2}{M_A^2} +4\,\frac{M_H^2}{M_{H^\pm}^2}  \right) \right.\nn\\[2mm]
   &&~~~~~~~~~~~~~~~~-\tan\beta\,\left[\,6\left(1-\frac{M^2}{M_H^2}\right)
     \left(1+\ln\frac{M_H^2}{m_t^2}\right)
     -\left(1-\frac{M_H^2}{M_A^2}\right)\left(1-5\,\ln\frac{M_A^2}{m_t^2}\right)
     \right.\nn\\[2mm]
     &&~~~~~~~~~~~~~~~~~~~~~~~~~~~~\left.\left.
     +4\left(1-\frac{M_H^2}{M_{H^\pm}^2}\right)\,\ln\frac{M_{H^\pm}^2}{m_t^2}
     \,\right]
   \,+\,{\cal O}\left(\frac{m_t^2}{M_\Phi^2}\right)\,\right\}~,\\[2mm]
   \label{eq:FBox2l}
   F_\Box^{2\ell,\,{\scriptscriptstyle {\rm 1PI}}} &=& 
   F_\Delta^{h,\,2\ell,\,{\scriptscriptstyle {\rm 1PI}}}\nn\\
   &&-\,\frac{m_t^2\,\cot^2\beta}{24\,\pi^2\,v^2}\,\left[\,
    3\left(1-\frac{M^2}{M_H^2}\right)^2
    \!-5\left(1-\frac{M^2}{M_A^2}\right)^2
    \!-4\left(1-\frac{M^2}{M_{H^\pm}^2}\right)^2
    \,+\,{\cal O}\left(\frac{m_t^2}{M_\Phi^2}\right)\,\right]~.\nn\\
   \eea
   
Keeping into account that $m_t^2 = y_t^2 \,v^2\sin^2\beta/2$ and that
the BSM Higgs masses are of the form $M_\Phi^2 = M^2 +
\tilde\lambda\,v^2$, where $\tilde\lambda$ denotes generic
combinations of quartic couplings, inspection of
eqs.~(\ref{eq:FDh2l})--(\ref{eq:FBox2l}) reveals that the two-loop 1PI
contributions to the triangle and box form factors are all
proportional to $y_t^2\,$, times combinations of BSM Higgs masses that
scale like $\tilde\lambda\,v^2/M_\Phi^2\,$,
$\,\tilde\lambda\,v^2/M_\Phi^2\,\ln(M_\Phi^2/m_t^2)\,$, or, in the
case of the contributions from ``ladder-like'' box diagrams in the
second line of eq.~(\ref{eq:FBox2l}),
$\tilde\lambda^2\,v^4/M_\Phi^4$. These combinations do not grow
linearly with $\tilde\lambda\,$ when the latter is large, in contrast
with the corrections to the trilinear Higgs couplings in
eqs.~(\ref{eq:2loophhh-deg})--(\ref{eq:2loophhH-ben}), where the
two-loop part is enhanced by $\tilde\lambda$ w.r.t.~the one-loop
part. We stress here how the dependence of the BSM Higgs masses on the
quartic couplings alters the superficial counting of powers of
$\tilde\lambda$ in a given diagram.

\bigskip

There are, however, additional contributions to the Higgs--gluon form
factors that, in principle, contribute to the cross section for Higgs
pair production at the same order in the scalar couplings as the
two-loop corrections to $\lambda_{hhh}$ and $\lambda_{hhH}$. These
stem from WFR insertions on the Higgs legs and from the
renormalization of the relation between $v$ and the Fermi constant
$G_\mu$ that is factored out of the cross section in
eq.~(\ref{eq:partxs}).  We find:
\beq
F_\Delta^{h,\,2\ell}~=~-\frac{\delta v}{v}\,F_\Delta^{h,\,1\ell}~,~~~~~
F_\Delta^{H,\,2\ell}~=~-\frac{\delta v}{v}\,F_\Delta^{H,\,1\ell}~,~~~~~
F_\Box^{2\ell} ~=~ \left(\delta Z_{hh}-2\,\frac{\delta v}{v}\right)\,F_\Box^{1\ell}~,
\label{eq:FDelBox}
\eeq
where the one-loop form factors at vanishing external momentum are
given in section~\ref{sec:setup}, and explicit formulas for $\delta
Z_{hh}$ and $\delta v$ are given in eqs.~(\ref{eq:dZhh}) and
(\ref{eq:dv}), respectively. The difference between the coefficients
of $\delta v/v$ in the triangle and box form factors stems from the
$v$ introduced in eq.~(\ref{eq:Cfactors}) to normalize $C_\Delta^h$
and $C_\Delta^H$.  Also note that we include a WFR counterterm only in
the box form factor because, in the $gg\rightarrow hh$ amplitude, the
Higgs legs of the triangle form factors correspond to the internal
$s$-channel propagators.

\section{Numerical impact of the two-loop BSM corrections}
\label{sec:numerical}

Before studying the impact of the two-loop BSM corrections on the
cross section for the production of a pair of SM-like Higgs bosons in
the aligned THDM, we illustrate their effect on the trilinear Higgs
couplings. For the coupling between three SM-like Higgs bosons, we
consider the modifier $\kappa_\lambda \equiv
\lambda_{hhh}\,/\,\lambda^{\smallSM}_{hhh}\,$, which is often
introduced in phenomenological discussions of BSM models. In the
alignment limit, and for the purposes of the present discussion, we
can express the coupling modifier as $\kappa_\lambda = (\lambda^{\rm
  tree}_{hhh} \,+\,\delta_0^{1\ell}\lambda_{hhh} \,+\,
\delta_0^{2\ell}\hat\lambda_{hhh})\,/\,\lambda^{\rm tree}_{hhh}\,$, where
$\lambda^{\rm tree}_{hhh} = 3\,M_h^2/v^2$ while the one- and two-loop
contributions at vanishing external momenta are those computed in
section~\ref{sec:lamhhh}.

\begin{figure}[t]
\begin{center}
  \vspace*{-1.3cm}
  \includegraphics[width=8.3cm]{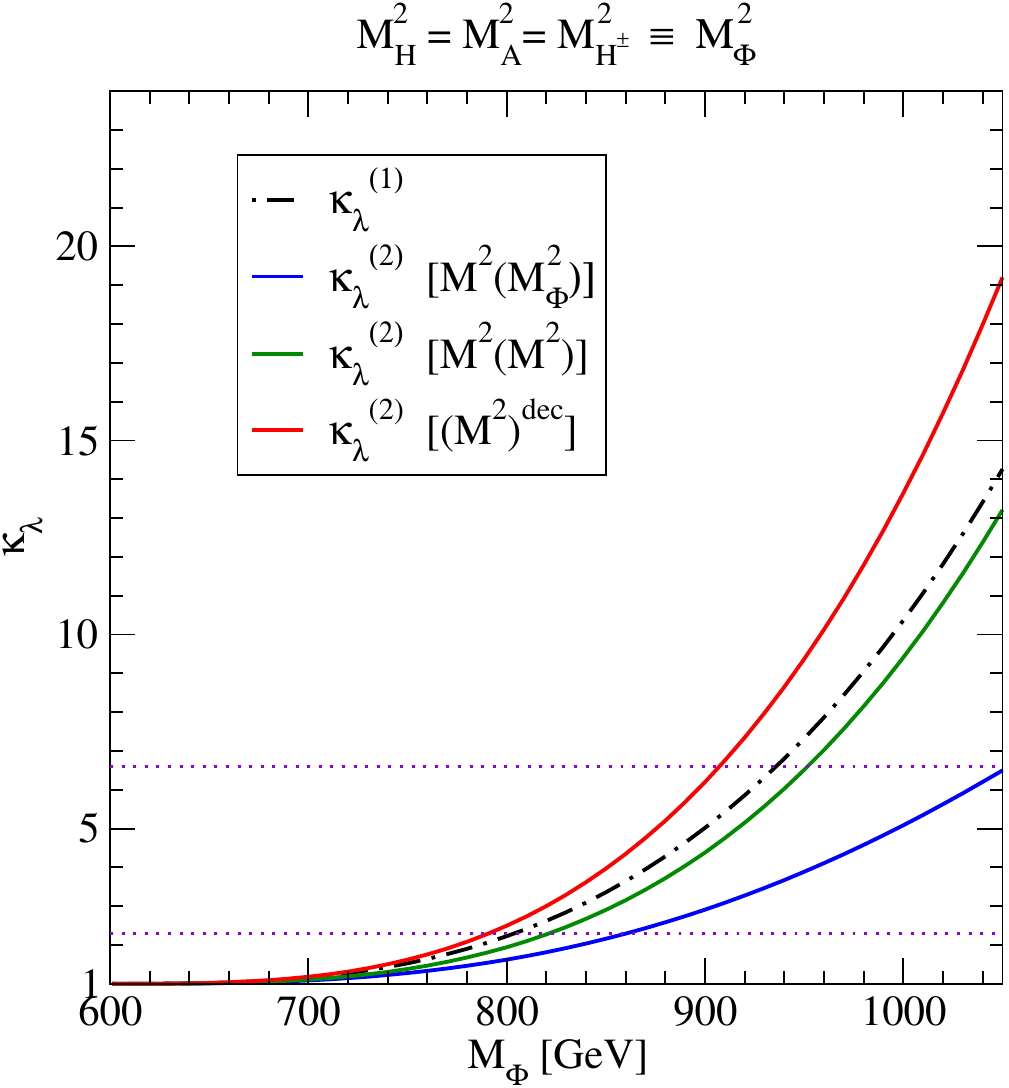}~~~
  \includegraphics[width=8.3cm]{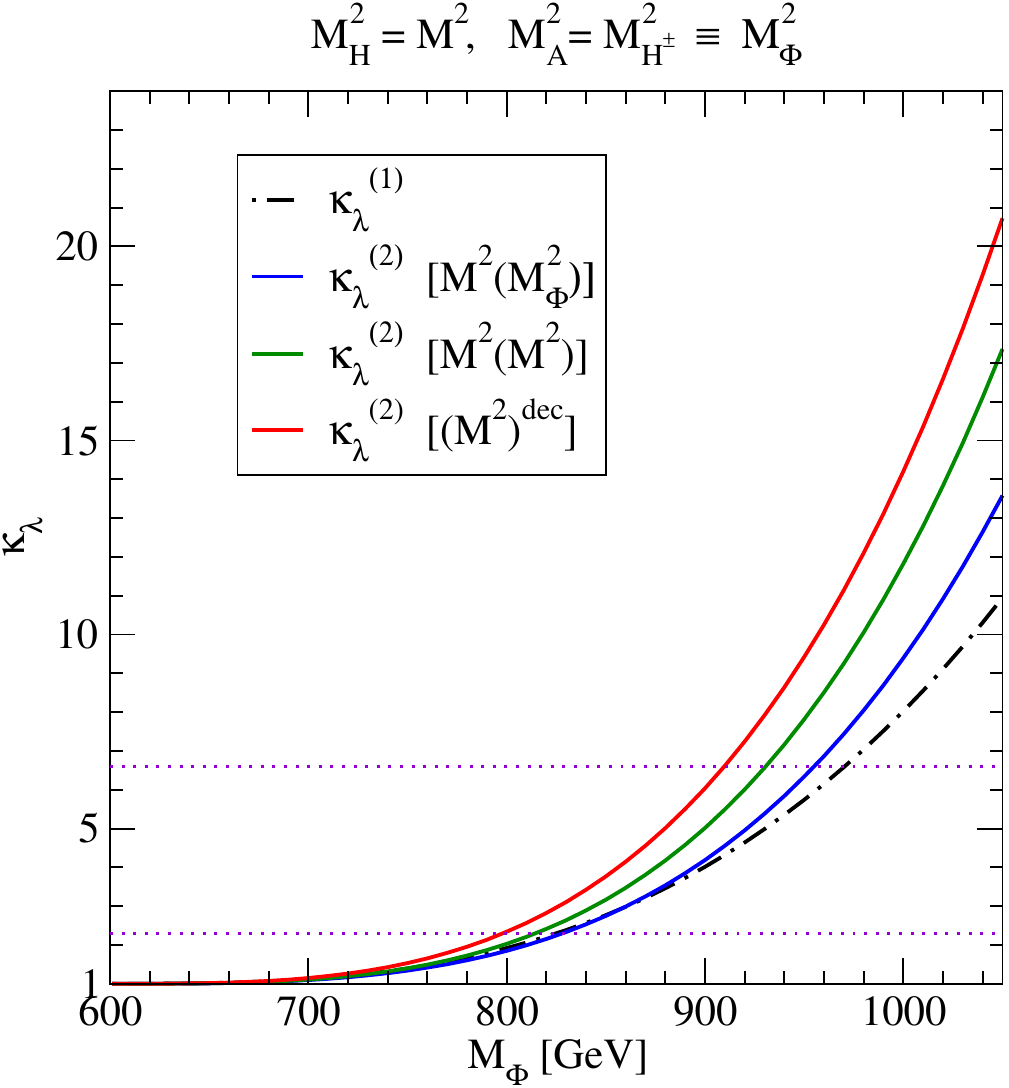}
  \caption{\em Left:~~Trilinear coupling modifier $\kappa_\lambda$ as
    function of a common pole mass $M_\Phi$ for the BSM Higgs bosons,
    in a benchmark scenario for the aligned THDM where $\MM=(600~{\rm
      GeV})^2$ and $\tan\beta=1.2$. The dot-dashed line is the one-loop
    result and the solid lines are two-loop results. The latter
    correspond to three different definitions of the parameter $\MM$,
    as explained in the text. Right:~~Same as in the left plot, for
    the scenario with $M_H^2 = \MM=(600~{\rm GeV})^2$ and
    $M_A^2=M_{H^\pm}^2 \equiv M_\Phi^2$.}
  \label{fig:kaplam}
\vspace*{-6mm}
\end{center}
\end{figure}

In figure~\ref{fig:kaplam} we examine the dependence of
$\kappa_\lambda$ on the pole masses of the BSM Higgs bosons in the
simplified scenarios for the aligned THDM that we introduced in
section~\ref{sec:formulas}, with either $M_H^2 = M_A^2 = M_{H^\pm}^2
\equiv M_\Phi^2$ (left plot) or $M_H^2 = \MM$,~ $M_A^2 = M_{H^\pm}^2
\equiv M_\Phi^2$ (right plot).  We can thus rely on
eq.~(\ref{eq:2loophhh-deg}) or on eq.~(\ref{eq:2loophhh-ben}),
respectively, to compute the one- and two-loop corrections to
$\lambda_{hhh}$. However, to facilitate the comparison with earlier
calculations, we modify those equations as indicated in
section~\ref{sec:formulas} to ensure that all three of the $h$ legs
are renormalized in the OS scheme. In both plots we take $\MM =
(600~{\rm GeV})^2$, and in the left plot we also take $\tan\beta =
1.2$ (in contrast, as mentioned earlier, the BSM-Higgs contributions
to $\kappa_\lambda$ do not depend on $\tan\beta$ when $M_H^2 =
\MM$). Note that our choice for the value of $\MM$ implies that the
right plot corresponds to the benchmark scenario introduced in
ref.~\cite{Bahl:2022jnx}.
In both plots we show $\kappa_\lambda$ as a function of $M_\Phi$,
denoting by $\kappa_\lambda^{(1)}$ the quantity obtained including
only the one-loop part of the corrections (black dot-dashed line), and
by $\kappa_\lambda^{(2)}$ the quantity that includes both the one- and
two-loop parts (blue, green and red solid lines). The three solid
lines differ in the definition of the parameter $\MM$. In particular,
the blue line is obtained when $\MM$ is an $\msbar$-renormalized
parameter at the scale $Q^2 = M_\Phi^2$; the green line is obtained
when $\MM$ is an $\msbar$-renormalized parameter at the scale $Q^2 =
\MM$; the red line corresponds instead to the use of the
``decoupling'' mass parameter $(\MM)^{\rm dec}$, as defined in
eq.~(\ref{eq:deltaM2}). The horizontal dotted lines correspond to the
most recent upper bound from ATLAS~\cite{ATLAS:2025hhd},
$\kappa_\lambda^{\rm exp} < 6.6$, and the HL-LHC
projection~\cite{Cepeda:2019klc}, $\kappa_\lambda^{\rm exp} < 2.3$. In
these scenarios, larger values of $M_\Phi$ correspond to larger values
of the quartic Higgs couplings, some of which become of ${\cal O}(10)$
towards the right edge of each plot. However, we evaluated the
constraints on the couplings from tree-level perturbative unitarity
following eqs.~(7)--(11) and (18)--(23) of ref.~\cite{Arco:2020ucn},
and verified that they are satisfied up to $M_\Phi \approx 1000$~GeV
in the left plot, and over the whole range of $M_\Phi$ in the right
plot. We remark that choosing a value of $\tan\beta$ farther away from
unity (in either direction) would lower the tree-level
perturbative-unitarity bound on $M_\Phi$ in the left plot, but would
leave it unchanged in the right plot, where $M_H^2 = \MM$.

\begin{figure}[t]
\begin{center}
  \vspace*{-1.3cm}
  \includegraphics[width=8.3cm]{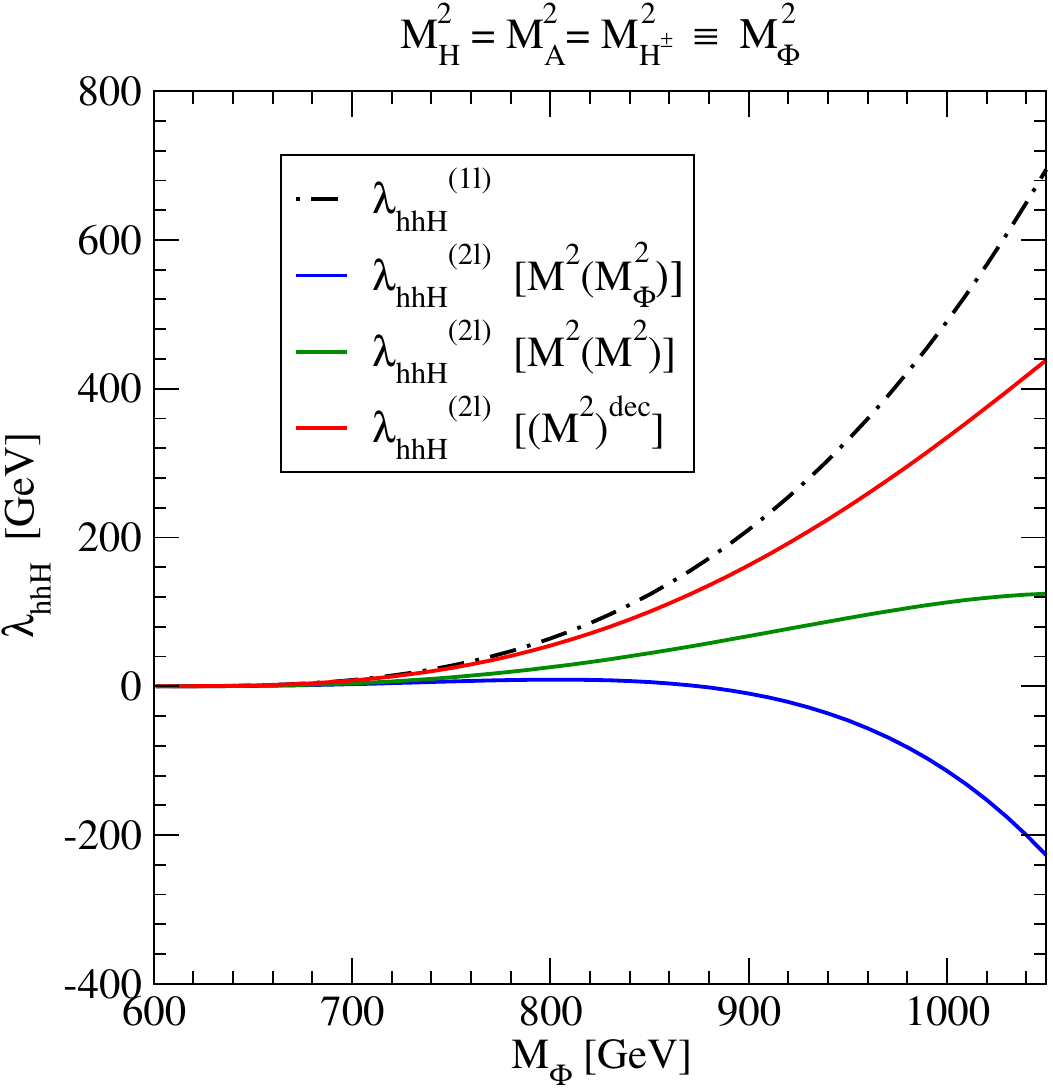}~~~
  \includegraphics[width=8.3cm]{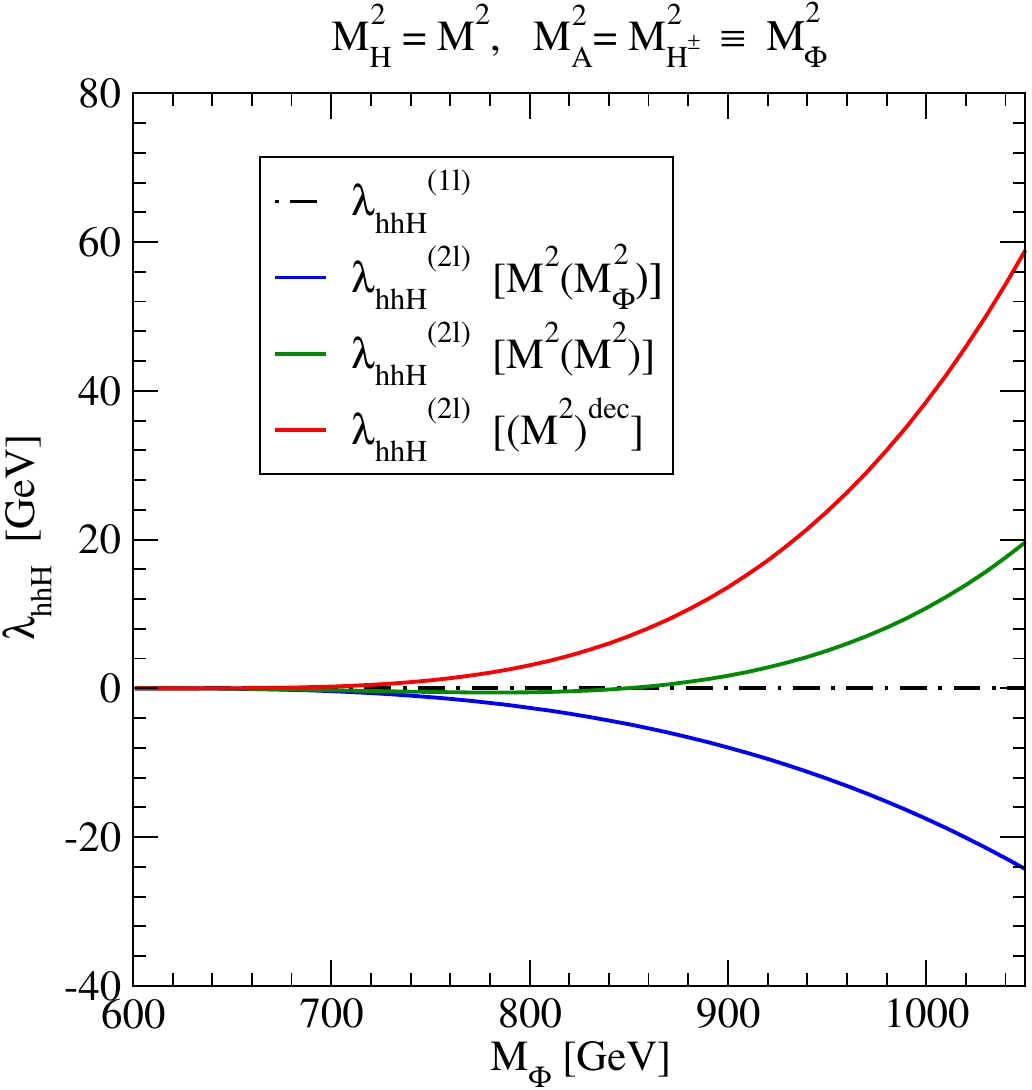}
  \caption{\em Left:~~Trilinear coupling $\lambda_{hhH}$ as
    function of a common pole mass $M_\Phi$ for the BSM Higgs bosons,
    in a benchmark scenario for the aligned THDM where $\MM=(600~{\rm
      GeV})^2$ and $\tan\beta=1.2$. The dot-dashed line is the
    one-loop result, the solid lines are two-loop results for three
    different definitions of the parameter $\MM$. Right:~~Same as in
    the left plot, for the scenario with $M_H^2 = \MM=(600~{\rm
      GeV})^2$, $M_A^2=M_{H^\pm}^2 \equiv M_\Phi^2$, and
    $\tan\beta=1.2$. Note the different scale on the $y$ axis
    w.r.t.~the left plot.}
  \label{fig:lamhhH}
\vspace*{-6mm}
\end{center}
\end{figure}

\bigskip
The comparison between the dot-dashed line and the solid lines in each of
the plots of figure~\ref{fig:kaplam} shows that the two-loop BSM-Higgs
corrections to the trilinear self-coupling of the SM-like Higgs boson
can be comparable in size with the one-loop corrections, and they can
have a significant impact on the bounds on the BSM-Higgs masses that
can be extracted from the comparison with the experimental bound on
$\kappa_\lambda$. However, the spread among the solid lines shows that
any statement on the significance (and, in the left plot, even the
direction) of the two-loop corrections is strongly dependent on the
definition adopted for the parameter $\MM$ entering the one-loop part
of the corrections. We stress that the three solid lines in each of
the plots of figure~\ref{fig:kaplam} map three different regions of
the THDM parameter space, thus their spread should not be interpreted
as a theory uncertainty of the two-loop calculation.
For example, for $M_\Phi = 1$~TeV and $(\MM)^{\rm dec} = (600~{\rm
  GeV})^2$, eq.~(\ref{eq:deltaM2}) implies that
$\MM(\MM)\approx(513~{\rm GeV})^2$ and $\MM(M_\Phi^2)\approx(406~{\rm
  GeV})^2$ in the left plot, and $\MM(\MM)\approx(536~{\rm GeV})^2$
and $\MM(M_\Phi^2)\approx(462~{\rm GeV})^2$ in the right plot.
This said, in scenarios where $\MM$ is kept fixed and the growth of
the BSM-Higgs masses is driven by the growth of the quartic Higgs
couplings -- hence, we cannot rely on decoupling considerations -- we
see no obvious criterion to choose a priori whether we should use
$\MM(M_\Phi^2)$, $\MM(\MM)$, or $(\MM)^{\rm dec}$ as input
parameter. This ambiguity in the size of the two-loop corrections
might just have to be regarded as an undesirable aspect of BSM
scenarios in which some couplings are very large.

Finally, the black dot-dashed and red solid lines in the right plot of
figure~\ref{fig:kaplam} can be compared with the one- and two-loop
lines in figure~2 of ref.~\cite{Bahl:2022jnx}, which relies on the
calculation of $\lambda_{hhh}^{2\ell}$ from
refs.~\cite{Braathen:2019pxr,Braathen:2019zoh}, with all three of the
$h$ legs renormalized in the OS scheme and $\MM$ interpreted as the
``decoupling'' mass parameter. We find good agreement between the
predictions for $\kappa_\lambda^{(1)}$, whereas our prediction for
$\kappa_\lambda^{(2)}$ is somewhat lower than the one in
ref.~\cite{Bahl:2022jnx}. We point the reader to
section~\ref{sec:lamhhh} for an explanation of the discrepancies
between our calculation of $\lambda_{hhh}^{2\ell}$ and the one of
refs.~\cite{Braathen:2019pxr,Braathen:2019zoh}.

\bigskip

In figure~\ref{fig:lamhhH} we illustrate the effect of the two-loop
BSM corrections on the coupling between two SM-like Higgs bosons and
one BSM Higgs boson, in the same benchmark scenarios for the aligned
THDM that we considered in figure~\ref{fig:kaplam}. In this case we do
not introduce a coupling modifier, but plot directly the coupling
$ \lambda_{hhH}$ that enters the amplitude for Higgs pair
production via eqs.~(\ref{eq:ChNNLO}) and (\ref{eq:CHNNLO}).  We
can thus rely directly on eq.~(\ref{eq:2loophhH-deg}), for the left
plot, and on eq.~(\ref{eq:2loophhH-ben}), for the right plot, to
compute the one- and two-loop corrections to $ \lambda_{hhH}$.  As
in figure~\ref{fig:kaplam}, the black dot-dashed line is obtained
including only the one-loop part of the corrections, while the three
solid lines are obtained including both the one- and two-loop parts,
and differ in the definition of the parameter $\MM$. In particular,
the blue, green, and red solid lines are obtained by requiring that
$\MM(M_\Phi^2)$, $\MM(\MM)$, and $(\MM)^{\rm dec}$, respectively, be
equal to $(600~{\rm GeV})^2$.

The comparison between the dot-dashed line and the solid lines in the
left plot of figure~\ref{fig:lamhhH} shows that, even in the case of $
\lambda_{hhH}$, the two-loop corrections can become comparable with
the one-loop corrections for large values of $M_\Phi$, where the
quartic Higgs couplings are in turn large. Again, we find that the
size of the two-loop corrections depends strongly on the definition
adopted for the parameter $\MM$ entering the one-loop part. The right
plot of figure~\ref{fig:lamhhH} is different, because the one-loop
correction to $ \lambda_{hhH}$ vanishes for $M_H^2 = \MM$, while a
residual two-loop correction is induced by counterterm
contributions. Indeed, we zoomed the $y$ axis in by a factor 10
w.r.t.~to the left plot in order to improve the readability of the
four lines. We note that eq.~(\ref{eq:2loophhH-ben}) implies that a
larger value of $\tan\beta$ would result in a larger two-loop
correction to $ \lambda_{hhH}$.
However, the $\tan\beta$-independence of the perturbative-unitarity
bounds that we obtain at the tree level in this scenario is likely to
be violated at higher orders, in analogy to what we find for the
vanishing of $ \lambda_{hhH}$. Away from $\tan\beta\approx 1$, a
more accurate determination of those bounds would be necessary to
ensure that the considered range of $M_H$ is not ruled out. Moreover,
in the amplitude for the process $gg\rightarrow H\rightarrow hh$ any
enhancement of $ \lambda_{hhH}$ due to a larger value of
$\tan\beta$ would be compensated for by a suppression of the $ggH$
form factor, see eq.~(\ref{eq:FDelNLO}).

\begin{figure}[p]
\begin{center}
  \vspace*{-15mm}
  \includegraphics[width=13cm]{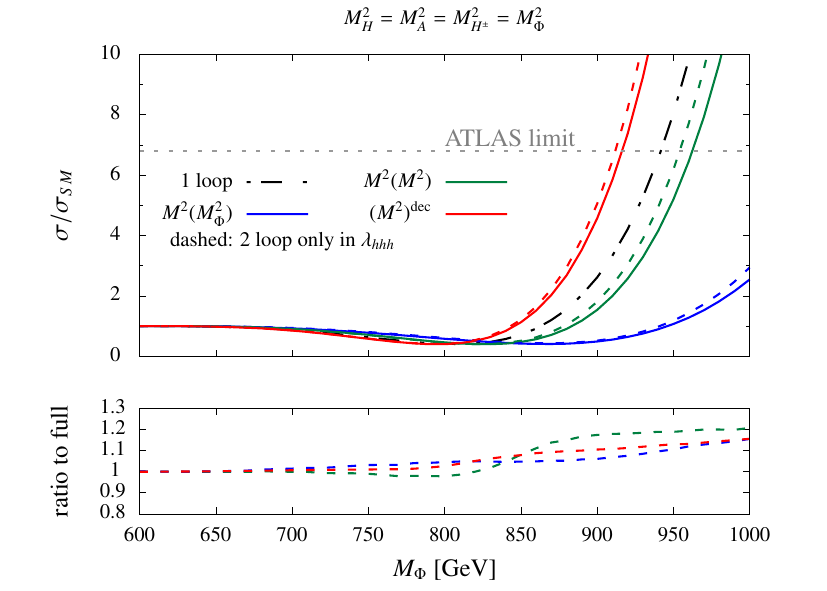}
  \caption{\em Top: Cross section for the production of a pair of
    SM-like Higgs bosons at the LHC with $\sqrt s = 13$~TeV,
    normalized to the SM prediction, as function of a common pole mass
    $M_\Phi$ for the BSM Higgs bosons, in a benchmark scenario for the
    aligned THDM where $\MM=(600~{\rm GeV})^2$ and
    $\tan\beta=1.2$. The meaning of the different lines is described
    in the text. Bottom: Ratios of the dashed lines over the solid
    lines.}
  \label{fig:XS1}
\vspace*{-5mm}
\end{center}
\end{figure}

\begin{figure}[p]
  \begin{center}
    \vspace*{5mm}
  \includegraphics[width=13cm]{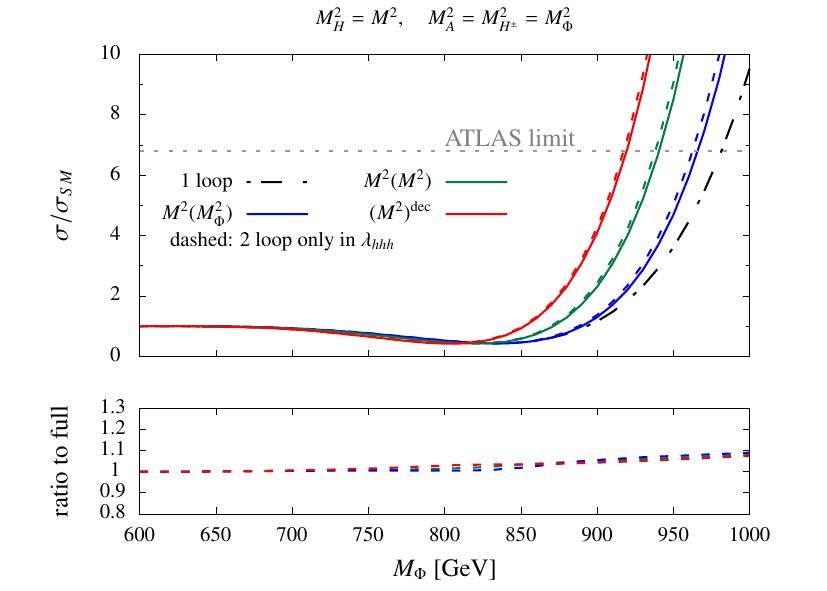}
  \caption{\em Same as figure~\ref{fig:XS1} for the scenario with
    $M_H^2 = \MM=(600~{\rm GeV})^2$ and $M_A^2=M_{H^\pm}^2 \equiv
    M_\Phi^2$.}
  \label{fig:XS2}
\vspace*{-5mm}
\end{center}
\end{figure}

\bigskip
We are now ready to discuss the combined impact of the two-loop BSM
corrections on the cross section for the production of a pair of
SM-like Higgs bosons in the aligned THDM. In figures \ref{fig:XS1} and
\ref{fig:XS2} we show the predictions for the cross section at the LHC
with $\sqrt s = 13$~TeV, normalized to the SM prediction computed at
the same order (i.e., the LO) in QCD, in the two benchmark scenarios
introduced earlier for the left and right plots of
figures~\ref{fig:kaplam} and \ref{fig:lamhhH}.  We used a modified
version of the public code {\tt HPAIR}~\cite{HPAIR} to compute the
pair-production cross section (setting
$\mu_\smallR=\mu_\smallF=M_{hh}/2$), and the code {\tt
  HDECAY}~\cite{Djouadi:1997yw, Djouadi:2018xqq} to compute the width
$\Gamma_H$ entering eqs.~(\ref{eq:ChNNLO}) and (\ref{eq:CHNNLO}).
In the upper plot of each figure, the black dot-dashed line represents
the prediction obtained at the NLO in the BSM Higgs couplings, i.e.,
using only eqs.~(\ref{eq:boxtriangle})--(\ref{eq:fV}) for the various
components of the spin-zero form factor ${\cal F}$. The blue, green
and red solid lines represent the full NNLO predictions, in which
${\cal F}$ is obtained from
eqs.~(\ref{eq:lambda2l})--(\ref{eq:CHNNLO}). The blue,
green and red dashed lines are instead obtained by including only the
two-loop corrections to $ \lambda_{hhh}$, and neglecting all of the
other sources of NNLO corrections.  As in the previous figures, the
blue, green and red lines (either solid or dashed) are obtained by
requiring that $\MM(M_\Phi^2)$, $\MM(\MM)$, and $(\MM)^{\rm dec}$,
respectively, be equal to $(600~{\rm GeV})^2$.  To illustrate the
collective impact of the NNLO corrections other than $\delta_0^{2\ell}
\lambda_{hhh}$, in the lower plot of each figure we show the ratio of
the blue, green and red dashed lines over the corresponding solid
lines. Finally, the grey dotted line at
$\sigma/\sigma_{\smallSM}\approx 6.8$ represents the upper bound on
the cross section that can be extracted from the most recent ATLAS
bound on $\kappa_\lambda$~\cite{ATLAS:2025hhd}.

The comparison between the black dot-dashed line and the solid lines
in figures~\ref{fig:XS1} and~\ref{fig:XS2} illustrates once again the
effect of the NNLO corrections controlled by the potentially large
BSM-Higgs couplings. In both scenarios, the qualitative dependence of
the prediction for the cross section on the definition of the
parameter $\MM$ (i.e., the relative position of the blue, green, and
red solid lines w.r.t.~the black dot-dashed line) matches the behavior
of the corrections to $\lambda_{hhh}$ shown in
figure~\ref{fig:kaplam}.  The comparison between the blue, green, and
red solid lines and the corresponding dashed lines in
figures~\ref{fig:XS1} and~\ref{fig:XS2} shows that, at least in these
scenarios, the two-loop corrections to $ \lambda_{hhh}$ make up the
bulk of the NNLO contributions to the cross section for Higgs pair
production. The remaining, sub-dominant contributions include the
two-loop corrections to $ \lambda_{hhH}$ in
eqs.~(\ref{eq:2loophhH-deg}) and (\ref{eq:2loophhH-ben}), the
corrections to the Higgs--gluon form factors in
eq.~(\ref{eq:FDelBox}), and the momentum-dependent corrections to the
$s$-channel propagators in eqs.~(\ref{eq:ChNNLO}) and
(\ref{eq:CHNNLO}). The ratios of dashed over solid lines in the lower
plot of figure~\ref{fig:XS1} show that, in the scenario where all BSM
Higgs masses are degenerate, the combined effect of the sub-dominant
NNLO contributions to the cross section can be up to about $20\%$ when
the couplings are large.
On the other hand, in the scenario of figure~\ref{fig:XS2}, where
$ \lambda_{hhH}$ is strongly suppressed (in particular, it vanishes at
one loop), the lower plot shows that the effect of the sub-dominant
NNLO contributions to the cross section remains within $10\%$.
In both scenarios, the apparent dominance of the corrections to
$ \lambda_{hhh}$ over the other sources of corrections validates our
use of an upper bound on $\sigma/\sigma_{\smallSM}$ derived from the
bound on $\kappa_\lambda$.

Finally we note that, even in the scenario of figure~\ref{fig:XS1}
where the loop-induced coupling $ \lambda_{hhH}$ can be sizeable (see
the left plot of figure~\ref{fig:lamhhH}), the contribution to the
pair-production cross section from diagrams with $s$-channel exchange
of $H$ is always within the $\pm 10\%$ range. This happens because at
large values of $M_\Phi$, where the BSM Higgs couplings are enhanced, the
resonant contribution to the $gg\rightarrow hh$ amplitude is
suppressed by a low gluon luminosity.

\section{Conclusions}
\label{sec:conclusions}
\vspace*{2mm}

The requirement that an extension of the SM accommodate a Higgs boson
with properties compatible with those observed at the LHC constrains
the parameter space of the BSM model even before the direct
observation of any new particles. However, the searches for Higgs pair
production at the LHC have not yet reached the sensitivity necessary
to test the SM prediction for the trilinear self-coupling of the Higgs
boson, and significant deviations from that prediction are in
principle still allowed.

In models with an extended Higgs sector, some of the quartic scalar
couplings can be significantly larger than any of the SM couplings,
while still satisfying all of the theoretical and experimental
constraints. In such cases, the radiative corrections controlled by
those couplings can be expected to be the dominant ones, and their
inclusion at the highest available order may prove necessary to obtain
accurate predictions for the properties of the observed Higgs
boson. In particular, it has long been known that, in the THDM, the
one-loop corrections involving BSM Higgs bosons can induce
modifications of ${\cal O}(100\%)$ to the trilinear self-coupling of
the SM-like Higgs boson~\cite{Kanemura:2002vm, Kanemura:2004mg}.

\bigskip

In this paper, we studied the impact of the radiative corrections
controlled by the BSM Higgs couplings on the cross section for the
production of a pair of SM-like Higgs bosons via gluon fusion in the
aligned THDM.  We relied on earlier work~\cite{Degrassi:2023eii} for a
definition of the alignment condition beyond the tree level. We
computed all of the contributions that affect the cross section up to
the NNLO in the potentially large scalar couplings, namely: the
two-loop corrections to the trilinear couplings $\lambda_{hhh}$ and
$\lambda_{hhH}$; the one-loop corrections to the $s$-channel
propagators entering the $gg\rightarrow hh$ amplitude; the two-loop
corrections to the Higgs-gluon form factors.  Since we are focusing on
scenarios where the BSM Higgs couplings are much larger than the SM
couplings, we worked at the LO in QCD, and we neglected the
corrections controlled by the EW-gauge and Yukawa couplings. We also
relied on the approximations of vanishing external momenta in the
two-loop diagrams and of vanishing mass for the SM-like Higgs boson.

Our calculation of the two-loop corrections to $\lambda_{hhh}$ allows
for a reassessment of the earlier result from
refs.~\cite{Braathen:2019pxr,Braathen:2019zoh}, and we found that the
latter omits contributions arising from the corrections to the
relation between the $h\Phi\Phi$ couplings, with $\Phi=(H,A,H^\pm)$,
the masses $M_\Phi^2$, and the parameter $\MM$. In contrast, our
calculation of the two-loop corrections to $\lambda_{hhH}$ is entirely
new, and entails several complications when compared with the case of
$\lambda_{hhh}$\,: it cannot rely solely on effective-potential
methods; it requires a separate counterterm for the $H\Phi\Phi$
couplings; it involves contributions related to the renormalization of
the mixing between light and heavy fields that cannot be absorbed in a
shift of the parameters entering the one-loop part of the
coupling. For both $\lambda_{hhh}$ and $\lambda_{hhH}$, we provided
explicit analytic formulas valid in two simplified scenarios where
some of the BSM Higgs masses are taken equal to each other, but we
make the general results available on request in electronic form.


We discussed the numerical impact of the two-loop BSM contributions,
first on the individual couplings and then on the prediction for the
pair-production cross section, in two benchmark scenarios for the
aligned THDM, one of which was previously introduced in
ref.~\cite{Bahl:2022jnx}. In both of these scenarios the parameter
$\MM$ is kept fixed, so that an increase in the BSM Higgs masses
corresponds to an increase in the quartic scalar couplings. We find
that, in the regions of the parameter space where the quartic scalar
couplings get large (but still allowed by the perturbative unitarity
bounds), the two-loop corrections to the trilinear couplings become
comparable in size with the one-loop corrections, thus their inclusion
becomes crucial in order to obtain precise predictions. This finding
carries over to the predictions for the pair-production cross
section. However, by analyzing the impact of the individual NNLO
contributions to the cross section, we see that the two-loop
corrections to $\lambda_{hhh}$ play by far the largest role, while the
remaning NNLO contributions (namely, corrections to $\lambda_{hhH}$,
propagator corrections, and corrections to the Higgs-gluon form
factors) account for a $10\%$--$20\%$ effect at most.

Finally, a recurring feature of the plots in
section~\ref{sec:numerical} is the fact that the size and even the
sign of the two-loop corrections depend strongly on the definition
adopted for the parameter $\MM$ entering the one-loop part of the
corrections. While this does not imply by itself a large uncertainty
of our predictions -- after all, a fixed value of $\MM$ corresponds to
different regions of the parameter space when the definition of $\MM$
changes -- it illustrates how sensitive BSM scenarios with very large
couplings can be to choices of renormalization scheme for the
parameters of the underlying Lagrangian, including parameters that do
not necessarily allow for a direct interpretation in terms of physical
observables.

\bigskip

The work presented in this paper could of course be extended in
multiple directions. For example, alternative definitions for $\MM$
and even for $\tan\beta$ could be explored, aiming to connect all of
the input parameters of the aligned THDM to physical observables. The
alignment condition itself could possibly be defined in different
ways. Finally, our calculation could be generalized to the case of a
non-zero (but necessarily small) mixing between the SM and BSM Higgs
doublets. Even for what concerns the numerical analysis in
section~\ref{sec:numerical}, our results should be regarded as merely
illustrative. A first improvement in our prediction for the
pair-production cross section would consist in combining the
contributions up to NNLO in the scalar couplings with a
state-of-the-art calculation of the QCD contributions. A systematic
scan over the parameter space of the aligned THDM, taking into account
all of the theoretical and experimental constraints, might then reveal
other interesting phenomena, such as regions where the sub-dominant
NNLO effects are larger than what we found here. We leave all of this
for future work, hoping that the results presented in this paper will
help the collective effort to use the properties of the Higgs boson as
a probe of what lies beyond the SM.

\bigskip
\bigskip

\section*{Acknowledgments}
This work received funding by the INFN Iniziativa Specifica APINE and by the
University of Padua under the 2023 STARS Grants$@$Unipd programme (Acronym and
title of the project: HiggsPairs – Precise Theoretical Predictions for Higgs
pair production at the LHC). This work was also partially supported by the
Italian MUR Departments of Excellence grant 2023-2027 ``Quantum Frontiers''.
We acknowledge support by the COST Action COMETA\_CA22130.

\newpage

\section*{Appendix: Details on the renormalization of the Higgs sector}
\setcounter{equation}{0}
\renewcommand{\theequation}{A\arabic{equation}} In this appendix we
provide additional details on the renormalization of the Higgs sector
of the aligned THDM. We first summarize the renormalization of the
mixing between light and heavy fields in the Higgs basis, then we
discuss how to consistently connect the Higgs couplings entering the
one-loop part of the corrections to $\lambda_{hhh}$ and
$\lambda_{hhH}$ with the parameter $M^2 \equiv 2\,m_{12}^2/\sin2\beta$.

\paragraph{Mixing renormalization in the Higgs basis:}
The so-called Higgs basis of the THDM is the basis in which one of the
SU(2) doublets develops the full SM-like vev $v$, and the other has
vanishing vev:
\beq
\Phi_\smallSM ~=~ \left(\!\begin{array}{c} G^+ \\ \frac{1}{\sqrt2}
(v + \phi^0_\smallSM + i\,G^0)  \end{array}\!\right)~,~~~~~
\Phi_\smallBSM ~=~ \left(\!\begin{array}{c} H^+ \\ \frac{1}{\sqrt2}
(\phi^0_\smallBSM + i\,A)  \end{array}\!\right)~.
\label{eq:HB}
\eeq
The scalar potential in the Higgs basis reads
\bea
V_0 &=& M_{11}^2\, \Phi_\smallSM^\dagger \Phi_\smallSM
~+~ M_{22}^2 \,\Phi_\smallBSM^\dagger \Phi_\smallBSM
~-~ M_{12}^2\,\left(\Phi_\smallSM^\dagger \Phi_\smallBSM ~+~ {\rm h.c.}\right)
\nn\\[2mm]
&&
+~ \frac{\Lambda_1}{2}\,\left(\Phi_\smallSM^\dagger \Phi_\smallSM\right)^2
\,+~ \frac{\Lambda_2}{2}\,\left(\Phi_\smallBSM^\dagger \Phi_\smallBSM\right)^2
\nn\\[2mm]
&&
+~ \Lambda_3\,\left(\Phi_\smallSM^\dagger \Phi_\smallSM\right)
\left(\Phi_\smallBSM^\dagger \Phi_\smallBSM\right)
+~ \Lambda_4\,\left(\Phi_\smallSM^\dagger \Phi_\smallBSM\right)
\left(\Phi_\smallBSM^\dagger \Phi_\smallSM\right)\nn\\[2mm]
&&
+~ \left[\,
\frac{\Lambda_5}{2}\,\left(\Phi_\smallSM^\dagger \Phi_\smallBSM\right)^2
+\, \left( \Lambda_6 \,\Phi_\smallSM^\dagger \Phi_\smallSM
\,+\,\Lambda_7\,\Phi_\smallBSM^\dagger \Phi_\smallBSM \right)
\,
\Phi_\smallSM^\dagger \Phi_\smallBSM ~+~{\rm h.c.}\right]~.
\label{eq:Vhb}
\eea
The relations between the mass and coupling parameters in
eq.~(\ref{eq:Vhb}) and the analogous parameters of the basis
$(\Phi_1,\Phi_2)$ in which the $Z_2$ symmetry that forbids
flavor-changing neutral-current interactions is manifest are given,
e.g., in the appendix of ref.~\cite{Davidson:2005cw}. We focus now on
the terms in the scalar potential that determine the mixing between
light and heavy fields. At the one-loop level, the minimization of the
potential w.r.t.~the neutral scalar component of the BSM Higgs doublet
implies
\beq
M_{12}^2 ~=~ \frac{\Lambda_6}{2}\,v^2
~+~ \frac{T^{1\ell}_{\phi_\smallBSM^0}}{v}~.
\label{eq:minhbloop}
\eeq
Combining eqs.~(\ref{eq:HB}) and (\ref{eq:Vhb}), we find that the
potential contains the light-heavy mixing terms
\beq
V_0 ~\supset~ \left(-M_{12}^2 + \frac{3\Lambda_6}2\,v^2 \right)
\,\phi_\smallSM^0\phi_\smallBSM^0
~+~\left(-M_{12}^2+\frac{\Lambda_6}2\,v^2\right)\,\left(G^0A \,+\,
G^+H^-\,+\,G^-H^+\right)~.
\label{eq:mixterms}
\eeq
Thus, the $(1,2)$ element of the one-loop-corrected mass matrix for the
neutral scalar components of the doublets, evaluated at the external
momentum $p^2$, reads
\beq
{\cal M}_{12}^2(p^2)~=~ -M_{12}^2 \,+\,\frac{3\Lambda_6}2\,v^2
\,+\, \Pi^{1\ell}_{\phi_\smallSM^0\phi_\smallBSM^0}(p^2)~.
\label{eq:massmix}
\eeq
As discussed in ref.~\cite{Degrassi:2023eii}, we define the alignment
condition at the one-loop level by requiring that ${\cal
  M}_{12}^2(p^2)$ vanish for $p^2=0$. In combination with the minimum
condition on $M_{12}^2$, see eq.~(\ref{eq:minhbloop}), and taking into
account that in the alignment limit $\phi^0_\smallSM = h$ and
$\phi^0_\smallBSM = -H$, this leads to
\beq
\Lambda_6 ~=~ \frac1{v^2} \,\left(\Pi^{1\ell}_{hH}(0)
- \frac{T^{1\ell}_H}v \right)~,
\label{eq:Lam6}
\eeq
while the light-heavy mixing terms in eq.~(\ref{eq:mixterms}) can be
expressed as
\beq
V_0 ~\supset~ 
-\Pi^{1\ell}_{hH}(0) \, hH
~+~ \frac{T^{1\ell}_H}v\,\left(G^0A \,+\,
G^+H^-\,+\,G^-H^+\right)~.
\label{eq:mixloop}
\eeq

\bigskip

In the calculation of the two-loop corrections to the trilinear
coupling $\lambda_{hhH}$ we must include the contributions from
one-loop diagrams with one insertion of the one-loop-induced
interactions given in eqs.~(\ref{eq:Lam6}) and (\ref{eq:mixloop}). We
collect those contributions in the correction $\Delta_{\rm mix}
\lambda_{hhH}$ given in eq.~(\ref{eq:dmix}). In particular, the terms
in the first two lines of eq.~(\ref{eq:dmix}), where we also made use
of the identities $\Pi^{1\ell}_{G^0A}(0)=
\Pi^{1\ell}_{G^{\pm}H^{\pm}}(0) = T^{1\ell}_H/v$, stem from diagrams
with the insertion of a light-heavy mixing term from
eq.~(\ref{eq:mixloop}) in one of the internal scalar lines. The terms
in the third and fourth lines of eq.~(\ref{eq:dmix}) stem instead from
diagrams in which one of the interaction vertices is proportional to
the coupling $\Lambda_6$ given in eq.~(\ref{eq:Lam6}). Finally, we
remark that there are no contributions of this kind in the case of
$\lambda_{hhh}$, because the relevant one-loop diagrams would require
two insertions of one-loop-induced interactions, thus amounting to
three-loop effects.

\paragraph{Connecting the Higgs couplings with $M^2$:}
In phenomenological analyses of the THDM, it is customary to express
the couplings between Higgs bosons as differences of squared mass
parameters divided by appropriate powers of $v$, and to use $M^2$, a
parameter of the $Z_2$-symmetric basis, as an input instead of the
Higgs-basis parameter $M^2_{22}$\,. At the tree level, the combination
of the minimum conditions of the scalar potential, the alignment
condition, and the approximation $M_h^2\approx0$ implies that
$M^2_{22}=M^2$\,. However, this relation receives one-loop
corrections, which must be taken into account in the calculation of
the two-loop corrections to $\lambda_{hhh}$ and $\lambda_{hhH}$. In
particular, we can write $M^2_{22}=M^2+ (\delta
M^2)_{\rm{\scriptscriptstyle cpl}}$, where the two squared mass
parameters are renormalized in the $\msbar$ scheme, and the
correction, which was derived in section 2 of
ref.~\cite{Degrassi:2023eii}, is the one given in
eq.~(\ref{eq:M22toM2}). Moreover, the corrections to the tree-level
relations between Higgs couplings, Higgs masses, and $M^2$ depend on
whether the couplings under consideration are of the $h\Phi\Phi$ kind
or of the $H\Phi\Phi$ kind, with $\Phi = (H,\,A,\,H^\pm)$. In the
alignment limit, the two kinds of couplings can be expressed in terms
of parameters of the Higgs basis as
\beq
\lambda_{h\Phi\Phi} ~\propto~ \frac{M_\Phi^2-M_{22}^2}{v}~,~~~~~~~~~~~~~
\lambda_{H\Phi\Phi} ~\propto~ \Lambda_7\,v~,
\label{eq:coupHB}
\eeq
where the form of the $\lambda_{h\Phi\Phi}$ couplings reflects the
fact that they depend on the same combinations of quartic couplings
that enter the terms proportional to $v^2$ in the tree-level masses
$M_\Phi^2$. The form of the $\lambda_{H\Phi\Phi}$ couplings reflects
instead the fact that the only term in the scalar potential that can
induce a coupling between three BSM Higgs bosons is the one involving
three BSM doublets and one SM doublet, see
eq.~(\ref{eq:Vhb}). Expressing the $\lambda_{h\Phi\Phi}$ couplings
entering the one-loop part of the corrections to $\lambda_{hhh}$ and
$\lambda_{hhH}$ in terms of $M^2$ instead of $M_{22}^2$, both
interpreted as $\msbar$-renormalized parameters, leads to the two-loop
shifts given in the second term of the second line of
eqs.~(\ref{eq:dlam-2loop}) and (\ref{eq:dlamhhH-2loop}),
respectively. Similarly, the conversion of the BSM Higgs masses
$M_\Phi^2$ entering the expression for $\lambda_{h\Phi\Phi}$ in
eq.~(\ref{eq:coupHB}) from $\msbar$-renormalized parameters to pole
masses is included in the two-loop shifts given in the first term of
the second line of eqs.~(\ref{eq:dlam-2loop}) and
(\ref{eq:dlamhhH-2loop}).

\bigskip

To express $\lambda_{H\Phi\Phi}$ as well in terms of a difference of squared
mass parameters, we rely on the fact that, as discussed in
ref.~\cite{Degrassi:2023eii}, the $(2,2)$ element of the
one-loop-corrected mass matrix for the neutral scalar components of
the doublets in the Higgs basis can be written as
\beq
{\cal M}_{22}^2(p^2)~=~ M^2 ~+~ \wt{\Lambda}\,v^2 ~+~
\Pi^{1\ell}_{\phi^0_\smallBSM\phi^0_\smallBSM}(p^2) ~-~\frac1v\,\left(
T^{1\ell}_{\phi^0_\smallSM} + 2\,\cot2\beta\,T^{1\ell}_{\phi^0_\smallBSM}\right)~,
\label{eq:MH22}
\eeq
where $\wt{\Lambda}$ is a combination of quartic couplings that, in
the case of the $Z_2$-symmetric THDM, can also be written as
$\wt{\Lambda} = (\Lambda_7 - \Lambda_6)\,\tan2\beta\,/2\,$. In the
alignment limit we can identify ${\rm Re}\,{\cal M}_{22}^2(M_H^2)$ with the pole
mass $M_H^2$, and invert eq.~(\ref{eq:MH22}) to obtain
\bea
\Lambda_7\!\!&=&\!\! \frac{2\,\cot2\beta}{v^2}\left[M_H^2-M^2
- {\rm Re}\,\Pi^{1\ell}_{HH}(M_H^2) +\frac12\,\tan2\beta\left(
\Pi^{1\ell}_{hH}(0)-\frac{T^{1\ell}_{H}}v\right)
+\frac1v\left(
T^{1\ell}_{h} - 2\,\cot2\beta\,T^{1\ell}_H\right)\right],\nonumber\\[1mm]
\label{eq:Lam7}
\eea
where we made use of the alignment condition on $\Lambda_6$, see
eq.~(\ref{eq:Lam6}). When the $\lambda_{H\Phi\Phi}$ couplings entering
the one-loop part of the corrections to $\lambda_{hhH}$ are expressed
as multiples of $2\,\cot2\beta\,(M_H^2-M^2)/v$, the term involving
${\rm Re}\,\Pi^{1\ell}_{HH}(M_H^2)$ in eq.~(\ref{eq:Lam7}) above is
included in the two-loop shifts given in the first term of the second
line of eq.~(\ref{eq:dlamhhH-2loop}). The remaining one-loop
corrections in eq.~(\ref{eq:Lam7}) are treated as a shift to the
parameter $M^2$, see eq.~(\ref{eq:MtildetoM2}), leading to the
two-loop shifts given in the third term of the second line of
eq.~(\ref{eq:dlamhhH-2loop}). We stress again that, in section
\ref{sec:lamhhH}, the parameter $\tilde M^2$ should be viewed merely
as a book-keeping device introduced to distinguish
$\lambda_{H\Phi\Phi}$ from $\lambda_{h\Phi\Phi}$ in
eqs.~(\ref{eq:lamhhH1lzeromom}) and (\ref{eq:dlamhhH-2loop}), and
that, after accounting for the different counterterm contributions to
those couplings, we can ultimately identify $\tilde M^2$ with $M^2$.

\newpage

\bibliographystyle{utphys} \bibliography{pair}

\end{document}